\documentclass[apj]{emulateapj}

\usepackage{lscape}
\usepackage{apjfonts}

\begin{document}

\newcommand{\Swift}{\textit{Swift}}
\newcommand{\Hete}{\textit{Hete-2}}
\newcommand{\Integral}{\textit{Integral}}
\newcommand{\Bepposax}{\textit{BeppoSAX}}
\newcommand{\KW}{\textit{Konus-Wind}}
\newcommand{\VLA}{\textit{Very Large Array}}
\newcommand{\HET}{\textit{Hobby-Eberly Telescope}}
\newcommand{\HST}{\textit{Hubble Space Telescope}}
\newcommand{\BATSE}{\textit{BATSE}}
\newcommand{\raptor}{\textit{RAPTOR}}

\title{Multi-Wavelength Observations of GRB\,050820A: An Exceptionally 
Energetic Event Followed from Start to Finish}

\author{S.~B.~Cenko\altaffilmark{1}, M.~Kasliwal\altaffilmark{2},
        F.~A.~Harrison\altaffilmark{1}, V.~Pal'shin\altaffilmark{3},
        D.~A.~Frail\altaffilmark{4}, P.~B.~Cameron\altaffilmark{2},
        E.~Berger\altaffilmark{5,6,7}, D.~B.~Fox\altaffilmark{8},
        A.~Gal-Yam\altaffilmark{2,7}, S.~R.~Kulkarni\altaffilmark{2},
        D.-S.~Moon\altaffilmark{1,9}, E.~Nakar\altaffilmark{10},
        E.~O.~Ofek\altaffilmark{2}, B.~E.~Penprase\altaffilmark{11},
        P.~A.~Price\altaffilmark{12}, R.~Sari\altaffilmark{10},
        B.~P.~Schmidt\altaffilmark{13}, A.~M.~Soderberg\altaffilmark{2},
        R.~Aptekar\altaffilmark{3}, D.~Frederiks\altaffilmark{3},
        S.~Golenetskii\altaffilmark{3}, D.~N.~Burrows\altaffilmark{8},
        R.~A.~Chevalier\altaffilmark{14}, N.~Gehrels\altaffilmark{15},
        P.~J.~McCarthy\altaffilmark{5}, J.~A.~Nousek\altaffilmark{8},
        S.~E.~Persson\altaffilmark{5}, T.~Piran\altaffilmark{16}}

\altaffiltext{1}{Space Radiation Laboratory, 220-47, 
	California Institute of Technology, Pasadena, CA 91125}
\altaffiltext{2}{Division of Physics, Mathematics, and Astronomy, 105-24,
	California Institute of Technology, Pasadena, CA 91125}
\altaffiltext{3}{Ioffe Physico-Technical Institute, 26 Polytekhnicheskaya, 
	St Petersburg 194021, Russian Federation}
\altaffiltext{4}{National Radio Astronomy Observatory, Socorro, NM 87801}
\altaffiltext{5}{Observatories of the Carnegie Institute of Washington,
	813 Santa Barbara Street, Pasadena, CA 91101}
\altaffiltext{6}{Princeton University Observatory, Peyton Hall, Ivy Lane,
	Princeton, NJ 08544}
\altaffiltext{7}{Hubble Fellow}
\altaffiltext{8}{Department of Astronomy and Astrophysics, Pennsylvania
	State University, 525 Davey Laboratory, University Park, PA 16802}
\altaffiltext{9}{Robert A.~Millikan Fellow}
\altaffiltext{10}{Theoretical Astrophysics, California Institute of Technology,
	130-33, Pasadena, CA 91125}
\altaffiltext{11}{Pomona College Department of Physics and Astronomy, 610
	N. College Ave, Claremont, CA 91711}
\altaffiltext{12}{Institute for Astronomy, University of Hawaii, 2680 Woodlawn
	Drive, Honolulu, HI 96822}
\altaffiltext{13}{Research School of Astronomy and Astrophysics, Australian
	National University, Mt Stromlo Observatory, via Cotter Rd, Weston
	Creek, ACT 2611, Australia}
\altaffiltext{14}{Department of Astronomy, University of Virginia, P.O. Box 
	3818, Charlottesville, VA 22903}
\altaffiltext{15}{NASA Goddard Space Flight Center, Greenbelt, MD 20771}
\altaffiltext{16}{The Racah Institute of Physics, Hebrew University, 
	Jerusalem 91904, Israel}

\email{cenko@srl.caltech.edu}

\shorttitle{Multi-Wavelength Observations of GRB\,050820A}
\shortauthors{Cenko et al.}

\begin{abstract}

We present observations of the unusually bright and long $\gamma$-ray
burst GRB\,050820A, one of the best-sampled broadband data sets in the \Swift\
era.  The $\gamma$-ray light curve is marked by a soft precursor pulse some 200
s before the main event; the lack of any  
intervening emission suggests that it is due to a physical mechanism
distinct from the GRB itself. The large time lag between the precursor and 
the main emission enabled simultaneous observations in the $\gamma$-ray,
X-ray, and optical band-passes, something only achieved for a handful
of events to date.  While the contemporaneous
X-rays are the low-energy tail of the 
prompt emission, the optical does not directly
track the $\gamma$-ray flux.  Instead,
the early-time optical data 
appear mostly consistent with the forward shock
synchrotron peak passing through the optical, and are therefore likely
the beginning of the afterglow.  On hour time scales after the burst, the
X-ray and optical light curves are 
inconsistent with an adiabatic expansion of the
shock into the surrounding region, but rather indicate that there is
a period of energy injection.  Observations at late times allow us to
constrain the collimation angle of the relativistic outflow to
$6.8^{\circ} \lesssim \theta \lesssim 9.3^{\circ}$.  Our
estimates of both the kinetic energy of the afterglow ($E_{\mathrm{KE}} = 
5.2^{+7.9}_{-4.1} \times 10^{51}$ ergs) and the prompt $\gamma$-ray
energy release ($E_{\gamma} = 7.5^{+6.7}_{-2.4} \times 10^{51}$ ergs)
make GRB\,050820A one of the most energetic events for which such values
could be determined. 

\end{abstract}



\section{Introduction}
\label{sec:intro}

With the discovery of the cosmological nature of $\gamma$-ray bursts (GRBs)
in 1997 
\citep{mdk+97}, astronomers were suddenly forced to explain the 
enormous isotropic energy release of these distant explosions.
Some of the most energetic events, such as 
GRB\,990123, seemingly released enough energy in the prompt $\gamma$-rays
($E_{\gamma,\mathrm{iso}} = 1.2 \times 10^{54}$ ergs; \citealt{bbk+99})
to rival the rest mass of a neutron star.  Furthermore, broadband
modeling of the best sampled events has shown that a comparable amount
of energy remains in the shock, powering the long-lived X-ray, optical,
and radio afterglow (see e.g.~\citealt{pk01,yhs+03}). 

The hypothesis that GRBs are a-spherical explosions \citep{r99}, 
supported by the appearance of achromatic ``jet'' breaks in a large
number of afterglow light curves \citep{sph99}, proved to be a turning 
point.  With typical opening angles of a few degrees, the true energy
release from most GRBs is $\sim 10^{51}$ ergs, on par with that of a
supernova (SN).  This realization enabled
the discovery of a standard energy reservoir for the collimation-corrected
prompt energy \citep{fks+01} and kinetic energy of the afterglow
\citep{bkf03}.  GRBs are now considered promising standard candle candidates,
with the hope of Hubble diagrams out to 
$z \approx 6$ offering complementary constraints to Type Ia SNe
on the cosmology of our universe (\citealt{fag+06,dlx04}; 
c.f.~\citealt{fb05}).  

Launched in November 2004, the \Swift\ Gamma-Ray Burst Explorer \citep{gcg+04}
was designed to position GRBs, disseminate accurate coordinates
to ground-based observatories in real-time, and follow the UV and X-ray
afterglows from minutes to days after the event.  In only a year
of full operation, \Swift\ has brought about
a number of fundamental advances in the GRB field, 
including the discovery of the first X-ray
(GRB\,050509b; \citealt{gso+05}) and near-infrared (GRB\,050724; 
\citealt{bpc+05}) afterglows
of a short-hard burst, 
the detection of the high-redshift ($z = 6.3$)
GRB\,050904 \citep{hnr+06,cmc+06}, and the ability to measure broadband
light curves starting shortly after, and in a few cases even during the
$\gamma$-ray event itself.

Despite these advances, measuring the bolometric fluences of \Swift\ events
has proved challenging, for a number of reasons.  First, the limited
energy range of the \Swift\ Burst Alert Telescope (BAT; \citealt{bbc+05})
means that \Swift\ can accurately characterize only the softest GRB spectra.
Second, few \Swift\ events have shown conclusive signs of a jet break, leaving
geometric corrections highly uncertain.  Finally, the X-ray light curves
of \Swift\ afterglows have shown both bright flares \citep{brf+05}
and slow decays \citep{nkg+06}.  Both behaviors have been attributed to
late-time ($t >> t_{\mathrm{GRB}}$) energy injection, and at times
have rivaled the energy release of the prompt $\gamma$-ray emission
(see e.g.~\citealt{fbl+06}).

On 20 August 2005 UT, the BAT detected and localized the unusually bright
and long GRB\,050820A, a truly rare burst in the \Swift\ sample.  The 
$\gamma$-ray light curve is marked by a soft pulse of emission preceding the 
main event by over 200 s.  The main emission was bright enough to be
detected by the \KW\ instrument, providing a $\gamma$-ray spectrum 
extending beyond 1 MeV as well as continuous coverage over the entire $\sim$
600 s burst duration.

Since \Swift\ triggered on the precursor, both space- and ground-based
facilities were able to image the transient during the bulk of the prompt
emission.  Such contemporaneous multi-wavelength observations have only been
achieved for a handful of bursts to date.  The bright X-ray ($F_{\nu} \sim
0.7$ mJy) and optical ($R \sim 14.5$ mag) afterglows made it possible
to study the evolution of the afterglow for weeks after the burst, providing
one of the most detailed broadband light curves in the \Swift\ era.  Finally,
late-time \HST\ observations allowed us to constrain the jet-break time, 
and hence the geometry of the outflow.  Even after applying the collimation
correction, we find GRB\,050820A is an exceptionally energetic event.

Our work proceeds as follows: in \S\ref{sec:obs}, we outline our broadband 
observations of the GRB\,050820A, beginning with the high-energy prompt
emission and followed by the X-ray, optical, and radio afterglow.  We
find the afterglow data are incompatible with the standard model of synchrotron
radiation from a single, highly-relativistic shock expanding adiabatically
into the surrounding medium \citep{spn98}.  Instead in \S\ref{sec:anal} we
use power-law fits 
($F_{\nu} \propto t^{-\alpha} \nu^{-\beta}$) to model the afterglow, dividing
the burst into segments based on noticeable temporal breaks in the X-ray
and optical light curves.  This analysis allows us to investigate
the early broadband light curve (\S\ref{sec:pre} and \ref{sec:early}),
late-time ($t >> t_{GRB}$) energy injection
in the forward shock (\S\ref{sec:inject}),
the structure of the circum-burst medium (\S\ref{sec:cbm}), 
and the geometry and
energetics of the event (\S\ref{sec:energy}).

Throughout this work we adopt a standard cosmology with $H_{0} = 71$ km 
s$^{-1}$ Mpc$^{-1}$, $\Omega_{\mathrm{M}} = 0.73$, and 
$\Omega_{\Lambda} = 0.27$.  We also make use of the notation $Q_{X} \equiv 
10^{X} \times Q$.

\section{Observations and Data Reduction}
\label{sec:obs}

In this section, we present our broadband observations of GRB\,050820A,
which span the spectral range from $\gamma$-rays to radio frequencies and 
extend in time out to 61 days after the burst.  
We include an independent analysis of the \Swift\ BAT
data set, as well as the
complete light curve and spectrum from the \KW\ instrument
\citep{afg+95}, which,
unlike the BAT, was able to observe GRB\,050820A over its entire
duration (\S\ref{sec:gamma}).  In \S\ref{sec:xray} we provide an
analysis of the \Swift\ X-Ray Telescope (XRT; \citealt{bhn+05})
data, with afterglow detections out to two weeks
after the event.  We present contemporaneous optical data
from the automated Palomar 60-inch telescope (P60; Cenko et al.~2006, 
in preparation) and
the \Swift\ Ultra-Violet Optical Telescope (UVOT; \citealt{rkm+05}),
supplemented by late-time images taken with the 9.2-m
\HET\ (HET) and the \HST\ (HST) (\S\ref{sec:opt}).
Finally, we monitored GRB\,050820A in the radio with the Very Large Array
beginning only hours after the burst and continuing for
approximately two months (\S\ref{sec:radio}).

\begin{figure*}
\epsscale{1.0}
\plotone{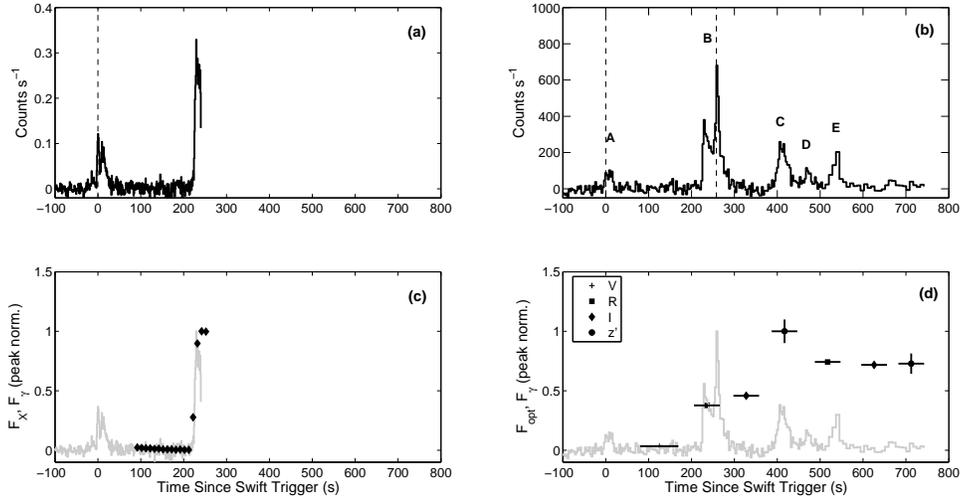}
\caption[Early broadband emission from GRB\,050820A]
        {Early Broadband Emission from GRB\,050820A.
        \textit{(a)}: \Swift-BAT light curve extracted from $15-350$ keV in 1 s
        bins.  The dashed
        vertical line is the time of \Swift\ trigger, 06:34:53 on 20 August
        2005 UT ($T_{\mathrm{BAT}}$).  While the
        second, brighter period of emission is clearly visible, \Swift\ entered
        the SAA approximately 240 s
        after $T_{\mathrm{BAT}}$, effectively terminating
        the observations.
        \textit{(b)}: \KW\ light curve extracted from $18-1150$ keV.
        The 5 peaks visible in the \KW\
        light curve are labeled $A-E$, and defined in Table \ref{tab:kwpeaks}.
        The left dashed vertical line shows
        the \Swift\ trigger time, while the right dashed line shows the \KW\
        trigger time, $T_{\mathrm{KW}}$, 258 s later.
        The portion
        of the light curve covered by the BAT comprises only a small
        fraction of the total $\gamma$-ray emission.
        \textit{(c)}: Contemporaneous \Swift\ XRT observations (black diamonds)
        overlaid on the BAT
        light curve.  The X-ray data nicely track the $\gamma$-ray emission.
        \textit{(d)}: Contemporaneous UVOT and P60 optical data
        overlaid on the \KW\ light curve.  Unlike the X-ray, the optical
        is not a good trace of $\gamma$-ray emission.  }
\label{fig:early_all}
\end{figure*}

\subsection{Gamma-ray Observations}
\label{sec:gamma}

\subsubsection{Swift Burst Alert Telescope}
\label{sec:BAT}
At 06:34:53 on 20 August 2005 UT\footnote{It is customary
to refer to the burst trigger time as $T_{0}$, for it is assumed to coincide
with the beginning of the prompt emission.  Given the unique nature of the
high-energy emission from GRB\,050820A, we undertake a more detailed study
to determine exactly when the prompt emission began (i.e.~$T_{0}$)
in \S\ref{sec:pre}.  Times measured with reference to the \Swift\ trigger
time will be referred to as $t_{\mathrm{BAT}}$ throughout the remainder of 
this work.}, 
the BAT
triggered and located GRB\,050820A (\Swift\ trigger 151207, 
\citealt{GCN.3830}).  
The initial location calculated
on-board was a 4\arcmin\ error circle centered at $\alpha = 22^{\mathrm{h}}
29^{\mathrm{m}}35\farcs9$, 
$\delta = +19^{\circ}11\arcmin14\farcs2$ (J2000.0).  
\citet{GCN.3835} describe a 
multi-peaked light curve ($t_{90} = 26 \pm 2$ s) with clear spectral evolution
(hard-to-soft) within each peak.

Following the report of additional high-energy emission
from the \KW\ instrument (\citealt{GCN.3852}; see \S\ref{sec:kw}), the 
BAT team re-analyzed their full light curve and found evidence of a much 
stronger, harder episode of emission from GRB\,050820A \citep{GCN.3858}.  
Unfortunately the satellite entered the South Atlantic Anomaly (SAA)
approximately 240 seconds after the burst trigger, and thus estimates of
the properties of this second phase are highly uncertain.  

Here we have independently analyzed the BAT data from GRB\,050820A.
We have extracted the $15-350$ keV light curve
in 1 s timing bins 
using software tools from the \Swift\ data analysis
package\footnote{Part of NASA's High Energy
Astrophysics Software package, see 
http://heasarc.gsfc.nasa.gov/docs/software/lheasoft.}.
The result is shown in Figure~\ref{fig:early_all}a.  
In addition, we have extracted spectra for the two periods of high energy
emission covered by the BAT data (peaks $A$ and $B$, see \S\ref{sec:kw}).
We then fit these spectra to a power-law distribution of energies
($\mathrm{d}N / \mathrm{d}E \propto E^{-\Gamma}$).  
We find evidence for strong spectral evolution between these two intervals,
as the second peak is significantly harder than the first.
The results of this analysis are shown in Table \ref{tab:BAT}.

\begin{deluxetable}{rcc}[b]
  \tabletypesize{\footnotesize}
  \tablecaption{Spectral Properties of BAT $\gamma$-Ray Emission}
  \tablecolumns{3}
  \tablewidth{0pc}
  \tablehead{\colhead{Time Interval} &
             \colhead{$\Gamma$}
             & \colhead{$\chi_{\mathrm{r}}^{2}$ / d.o.f.} \\
             \colhead{($t_{\mathrm{BAT}}$, s)} & &
            }
  \startdata
        $-17$ -- $22$ & $1.74 \pm 0.08$ & 1.07 / 75 \\
        $217$ -- $241$ & $1.07 \pm 0.06$ & 0.95 / 76 \\
  \enddata
  \tablecomments{Spectra were fit to a power-law model of the form
        $\mathrm{d}N / \mathrm{d}E \propto E^{-\Gamma}$.  Errors reported are
        90\% confidence limits.}
\label{tab:BAT}
\end{deluxetable}

\subsubsection{\KW}
\label{sec:kw}
The main part of GRB 050820A triggered \KW\ 
at $T_{\mathrm{KW}}$ = 06:39:14.512 UT, 257.948~s after
the BAT trigger (taking into account the 3.564 s propagation delay
from \Swift\ to \textit{Wind}). It was detected by the S2 detector
which observes the north ecliptic hemisphere; the incident angle was
63$\fdg$2.  Count rates are continuously recorded by \KW\ in three
energy bands: G1(18--70 keV), G2(70--300 keV), and G3(300--1150 keV).
Data collected in this ``waiting mode'' are acquired in 2.944 s timing
bins.  The time history recorded in the three energy ranges can
be considered a continuous three-channel spectrum.

Immediately following the \KW\ trigger, the instrument began simultaneously
collecting data in ``trigger'' mode, as well.  From $T_{\mathrm{KW}}$ to
$T_{\mathrm{KW}} + 491.776$ s, 64 spectra, each composed of 101 energy
channels ranging from 18 keV -- 14 MeV, were accumulated.  The time 
resolution of these ``trigger mode'' spectra varies from 64 ms -- 8.192 s
and is determined by an automated on-board algorithm based on count rate.  
Data were then processed using standard \KW\ analysis tools and spectra
were fit with XSPEC.

\begin{figure}
\epsscale{1.0}
\plotone{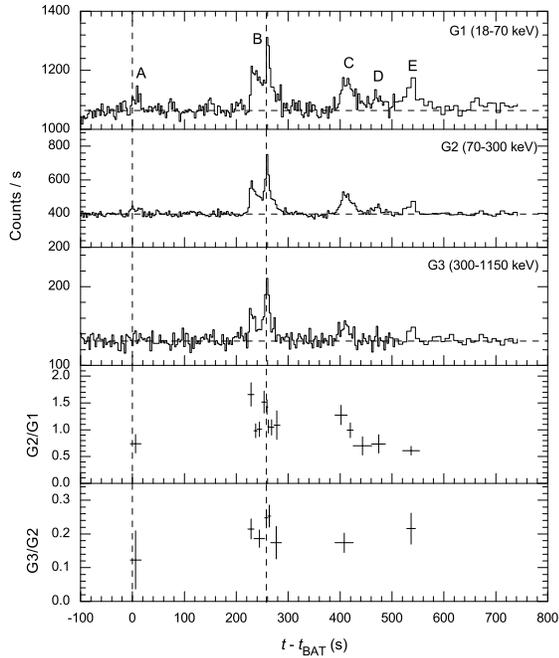}
\caption[Spectral evolution of GRB\,050820A]
        {Spectral evolution of GRB\,050820A.  The top three
        panels show the \KW\ light curve
        divided into three energy bands:  G1 $\equiv 18-70$ keV,
        G2 $\equiv 70-300$ keV, and G3 $\equiv 300-1050$ keV.
        The bottom two panels
        show the resulting hardness ratios: G2 / G1 and G3 / G2.
        Background levels are indicated with horizontal dashed lines.
        The vertical dashed lines denote the BAT ($T_{\mathrm{BAT}}$) and
        \KW\ ($T_{\mathrm{KW}}$) trigger times.
        With the notable exception of the precursor,
        the burst shows an overall hard-to-soft evolution, both over the entire
        duration and within individual bright peaks ($B$ and $C$).}
\label{fig:kwhr}
\end{figure}

At least five pulses are evident in the \KW\ light curve (labeled
$A-E$, see Fig.~\ref{fig:early_all}b and Table~\ref{tab:kwpeaks}).
Peak $A$, which generated the BAT trigger, appears to be a weak
precursor.
Figure~\ref{fig:kwhr} shows the three-channel light curve of
GRB\,050820A, as well as the hardness ratios. With the exception of
the precursor (peak $A$), the burst shows an overall hard-to-soft
evolution over the entire burst duration, as well as within some of
the individual peaks ($B$ and $C$).

The spectra of individual pulses are well fit by a cut-off power-law
model: $\mathrm{d}N / \mathrm{d}E 
\propto E^{-\alpha} \exp(-(2-\alpha)E/E_{\mathrm{p}})$; here
$\alpha$ is the photon index and $E_{\mathrm{p}}$ is the peak energy of the
$\nu \times F_\nu$ spectrum (see e.g.~\S\ref{sec:early}).  
Fitting the overall \KW\ ``trigger'' mode spectrum, accumulated from 
$0 < t_{\mathrm{KW}} < 295$ ($258 < t_{\mathrm{BAT}} < 553$)\footnote{After
this time only a weak tail is seen in the G1 band up to $t_{\mathrm{BAT}} 
\approx 730$ s; this tail contains less than 5\% of the total burst fluence.} 
in the 18--2000 keV range yields 
$\alpha=1.41_{-0.31}^{+0.25}$, and a peak energy $E_{\mathrm{p}} =
271_{-91}^{+359}$~keV (90\% c.l.; $\chi^2_{\mathrm{r}}=0.74$ for 62 d.o.f.).

However, in examining the full \KW\ light curve (Fig.~\ref{fig:early_all}b),
it is clear the above
time interval does not include a sizable fraction of the $\gamma$-ray
emission.  To
derive the spectral parameters of the time-integrated spectrum over
the main part of the GRB (peaks $B$--$E$), we simultaneously fit the
three-channel \KW\ spectrum accumulated from $-33 < t_{\mathrm{KW}} < 0$ 
and the overall multichannel spectrum. We find 
$\alpha=1.12_{-0.15}^{+0.13}$ and a peak energy $E_{\mathrm{p}} =
367_{-62}^{+95}$~keV (90\% c.l.; $\chi^2_{\mathrm{r}}=0.99$ for 64 d.o.f.).  
Not 
surprisingly the peak energy increased significantly, as the beginning of
Peak $B$ was the hardest portion of the entire burst.  We consider this
fit the most accurate estimate of the spectral properties of GRB\,050820A. 

The fluence and peak flux for each individual episode are shown in
Table~\ref{tab:kwpeaks}.  The total fluence received from
GRB\,050820A from 20--1000 keV (observer frame) was
$5.27^{+1.58}_{-0.69}
\times 10^{-5}$ ergs cm$^{-2}$ (90\% confidence limit).

\begin{deluxetable}{cccc}[!b]
  \tabletypesize{\footnotesize}
  \tablecaption{Properties of \KW\ $\gamma$-Ray Light Curve}
  \tablecolumns{4}
  \tablewidth{0pc}
  \tablehead{\colhead{Peak ID} & \colhead{Time Interval}
             & \colhead{Fluence\tablenotemark{a}}
              & \colhead{Peak Flux} \\
             & \colhead{($t_{\mathrm{BAT}}$, s)} &
             \colhead{($10^{-6}$ erg cm$^{-2}$)} &
              \colhead{($10^{-7}$ erg cm$^{-2}$ s$^{-1}$)}
            }
  \startdata
        $A$ & $-4.3$ -- $19.3$ & $2.77$ & $1.7$ \\
        $B$ & $222.4$ -- $282.8$ & $28.7$ & $13$ \\
        $C$ & $397.5$ -- $430.2$ & $10.2$ & $4.3$ \\
        $D$ & $454.8$ -- $479.4$ & $3.20$ & $1.9$ \\
        $E$ & $520.4$ -- $544.9$ & $4.93$ & $2.6$ \\
        \hline
        Total & $-4.3$ -- $544.9$ & $52.7^{+15.8}_{-6.9}$ & \\
  \enddata
  \tablenotetext{a}{The fluence was extracted from the 20--1000 keV energy
        range (observer frame) assuming a cut-off power-law spectrum
        of the form $\mathrm{d}N / \mathrm{d}E \propto E^{-\alpha}
        \exp^{-(2-\alpha)E / E_{\mathrm{p}}}$.  Errors reported are
        90\% confidence limits (see \S\ref{sec:kw} for
        further details).}
\label{tab:kwpeaks}
\end{deluxetable}

\subsection{X-ray Observations}
\label{sec:xray}

\label{sec:xrt}
The \Swift\ XRT began observations of GRB\,050820A 88
s after the burst trigger.  A bright, fading X-ray source was detected by
the automated on-board processing routine at $\alpha = 22^{\mathrm{h}} 
29^{\mathrm{m}} 37\farcs8$,
$\delta = +19^{\circ} 33' 32\farcs7$ (7\arcsec\ error radius)
and reported in real-time 
\citep{GCN.3830}.  Data was taken in Window Timing mode until $\sim 1.29$
hr  after the burst, when, due to the decreased count rate,
 Photon Counting mode was automatically initiated.

Here we have independently processed and reduced the XRT data from
GRB\,050820A (for a previous analysis of the early-time X-ray
data, see \citealt{owo+06}).  To reduce
the X-ray data, we used the software tools available from the {\em Swift} Data
Center.  We followed standard reduction steps, except for taking
measures to mitigate the effects of pulse pileup in our spectral
analysis.  During the steep rise in the X-ray light curve at $t_{\mathrm{BAT}}
\approx 220$ s, the XRT countrate exceeds the 200
counts~s$^{-1}$ at which pileup can affect the detector response
\citep{rcc+06}.  For
this segment, we removed the central two pixels from the point source
image.  At the beginning of Photon Counting mode, we removed the central four
pixels of the photon counting mode image for our spectral analysis,
again in order to mitigate the effect of pulse pileup \citep{vgb+06}.

We fit the X-ray spectrum for Phases 1a, 1b, and 2 separately (see 
\S\ref{sec:anal} for a full discussion of the division of the X-ray light
curve into phases).  We
used an absorbed powerlaw model, and fit both with the column fixed to
the Galactic value ($n_{\mathrm{H,Gal}} = 5.0 \times
10^{20}$~cm$^{-2}$; \citealt{dl90}), as well as with the column
floating.  For Phase 1a, the best fit column is consistent with
$n_{\mathrm{H,Gal}}$, and there is little difference between fits with
the column floating and fixed.  For both Phases 1b and 2 we find
acceptable fits with fixed column, however
the best fit columns are $1.5 \pm .23 \times 10^{21}$~cm$^{-2}$ and
$1.3 \pm .09 \times 10^{21}$~cm$^{-2}$ respectively.   Therefore we
find marginal evidence for absorption in excess of the Galactic
value (c.f.~\citealt{GCN.3837}).

The results of our analysis of the X-ray light curve of GRB\,050820A are
shown in Table~\ref{tab:xrt}.  
For the discussion below we adopt the spectral fits in Table~\ref{tab:xrayab}.
These were used to scale the binned count rates to fluxes and flux
density at a nominal energy of 5 keV.  

\begin{deluxetable*}{lrrcccc}
  \tabletypesize{\footnotesize}
  \tablecaption{X-ray Afterglow Spectral and Temporal Fits}
  \tablecolumns{9}
  \tablewidth{0pc}
  \tablehead{\colhead{Phase} & \colhead{$t^{\mathrm{start}}_{\mathrm{BAT}}$}
             & \colhead{$t^{\mathrm{stop}}_{\mathrm{BAT}}$} &
             \colhead{$\alpha$} &
             \colhead{$\chi^{2}_{\mathrm{r}}(\alpha)$ / d.o.f.}
             & \colhead{$\beta$} &
             \colhead{$\chi^{2}_{\mathrm{r}}(\beta)$ / d.o.f.} \\
             & \colhead{(s)} & \colhead{(s)} & & & &
            }
  \startdata
        1a & 0 & 217 & 2.2 $\pm$ 0.3 & 1.18 / 11 &
                $0.90 \pm 0.09$ & 0.63 / 349 \\
        1b & 217 & 257 & $\ldots$ & $\ldots$ &
                $-0.10 \pm 0.03$ & 1.02 / 749 \\
        2 & $4.8 \times 10^{3}$ & $8.7 \times 10^{4}$ & $0.93 \pm 0.03$ &
                1.14 / 29 & $1.20 \pm 0.04$ & 1.00 / 770 \\
        3 & $8.7 \times 10^{4}$ & $1.7 \times 10^{6}$ & $1.25 \pm 0.07$
                & $\ldots$ & $\ldots$ & $\ldots$ \\
  \enddata
  \tablecomments{We separately fit the X-ray light curves and spectra to
        power-law models of the form $F_{\nu} \propto t^{-\alpha}$ and
         $F_{\nu} \propto \nu^{-\beta}$, respectively.  The temporal decay
        in Phases 2 and 3 were fit jointly as a broken power-law, with
        the break time as a free parameter.  In Phase 3, we could not
        meaningfully constrain the spectral index due to the
        low count rate.  All errors quoted are 90\% confidence limits.}
\label{tab:xrayab}
\end{deluxetable*}

\subsection{Optical Observations}
\label{sec:opt}

\subsubsection{Palomar 60-inch Telescope}
\label{sec:p60}
The automated P60
responded to GRB\,050820A and began a pre-programmed
sequence of observations starting 3.43 minutes after the \Swift\
trigger.  The system is equipped with an optical CCD with a pixel scale
of 0.378\arcsec\ / pixel.  Images were taken in the Kron $R$ and $I$ 
and Gunn $g$ and $z$ filters.  All 
P60 images are processed with standard IRAF \citep{t86}
routines by an automated reduction pipeline in real-time.  Manual inspection of 
the first images revealed a bright variable
source inside the XRT error circle at $\alpha = 22^{\mathrm{h}} 
29^{\mathrm{m}} 38\farcs11$, $\delta = +19^{\circ} 33'37\farcs1$
(see Figure \ref{fig:finder}).
This position was promptly reported as the afterglow of 
GRB\,050820A 
\citep{GCN.3829}, allowing others to obtain high-resolution spectroscopy
of the afterglow \citep{GCN.3833}.  
We continued to monitor the
afterglow of GRB\,050820A with P60 for the following 7 nights, until it was
too faint for quantitative photometry.  

Optical photometry of the afterglow was complicated by the presence of two
nearby objects: one $R \sim 20.2$ mag star located $4.0$\arcsec\
south-west of the 
afterglow, and a fainter $R \sim 21.3$ mag object 
located only $2.9$\arcsec\ north-east
of the afterglow (see Figure \ref{fig:finder}, right panel).  
On some nights of poor seeing, the full-width at
half-maximum of our point-spread function (PSF) was 
larger than $2\farcs0$.  
We have therefore performed PSF-matched image subtraction
using the CPM 
technique of \citet{gmf+04} on our optical data.  Errors were
estimated by placing 5 artificial stars with flux equivalent to the afterglow
in locations with similar background contamination.  In addition, we have
also used aperture photometry (DAOPHOT) to extract the afterglow flux.  On the
first night, the afterglow was bright enough to be well-detected either in
single images or short co-additions ($\le 360$ s).  For these images,
both nearby sources were below our detection limit.  Results from
aperture photometry and image subtraction were therefore consistent.  We quote
our aperture photometry results for these data, as image defects
from imperfect PSF-subtraction seemed to artificially inflate these errors.
However, on subsequent nights the 
afterglow flux was either near or below the level of these nearby objects, and 
we therefore report results from our image subtraction technique. 

Photometric calibration was performed relative to 20 field stars
provided in \citet{GCN.3845}.  
Kron $R$ is similar to Cousins $R$ ($R_{\mathrm{C}}$) and was treated as identical for 
photometric calibration of these images.  Magnitudes from the standard
Johnson/Cousins system were transformed to Gunn $g$ using the empirical 
relation from \citet{g85}.  We found, however, a better correlation between
our $i$-band filter and 
the Cousins $I$ ($I_{\mathrm{C}}$)
 filter than from the transformation to Gunn $i$ provided in 
\citet{tg76}.  
We therefore use $I_{\mathrm{C}}$
in the remainder of this work.  For the Gunn $z$ filter, we used the 
Two Micron All Sky Survey (2MASS; \citealt{scs+06})
catalog and the optical photometry provided by \citet{GCN.3845} to interpolate
the spectral energy distributions of 10 sources to the Sloan $z'$
bandpass. 
Typical RMS variations in the calibration
sources were $0.03$ mag in $R_{\mathrm{C}}$ and $I_{\mathrm{C}}$, 
$0.04$ mag in $g$,
and $0.15$ in $z'$. 

The results of our P60 observations are shown in Table~\ref{tab:optical}.
While the expected galactic extinction in this direction is small 
($E(B - V) = 0.044$ mag; \citealt{sfd98}), we have incorporated it into our 
results because of the large wavelength range spanned by our observations
($E(U - z') = 0.17$ mag).  Errors quoted are 1$\sigma$
photometric and instrumental errors
summed in quadrature.  For this and all other optical data in this work,
magnitudes are converted to flux densities using the zero-points reported in 
\citet{fsi95}.

\subsubsection{\Swift\ Ultra-violet Optical Telescope}
\label{sec:uvot}
The \Swift\ UVOT automatically slewed 
to the BAT location and began observations only 80~s after the trigger.  
However, the UVOT also becomes inoperable in the SAA and therefore does not
cover the period of the main $\gamma$-ray emission.

The \Swift\ team reduced the early-time UVOT data and reported detections
in the $V$, $B$, $U$, and $UVW1$ bands \citep{GCN.3838}.  Here we have
independently reduced the $U$-, $B$-, and $V$-band
UVOT data following the recipe outlined in
\citet{ljf+06} (see \S3.6).  As a check, we re-calculated the $B$ and $V$ 
zero-points with respect to the field stars from \citet{GCN.3845}.  While 
our zero-points are consistent with the ones quoted in \citet{ljf+06}, 
we found a much larger scatter ($\sim 0.10$ mag  vs.~$0.01$ mag) that could
not be attributable solely to spread in the field stars.  We have therefore
incorporated the resulting zero-point errors for these
data points (as well as a similar value for $U$-band).
The results of these observations are shown in 
Table~\ref{tab:optical}.

\subsubsection{\HET}
\label{sec:het}
We triggered target-of-opportunity observations on the 9.2-m \HET\ beginning
on the night of 26 August UT.  Observations were taken in both $R_{\mathrm{C}}$ and $I_{\mathrm{C}}$
filters.  A second reference epoch was taken
on 29 August UT.  Image subtraction was performed on the two epochs to
remove contamination from nearby sources (as described in \S\ref{sec:p60}).
Photometric calibration was performed relative to 10 reference objects 
from \citet{GCN.3845}, and the absolute calibration was of similar quality
to the P60 dataset.  Our results are reported in Table \ref{tab:optical}.

\begin{figure*}
\epsscale{1.0}
\plotone{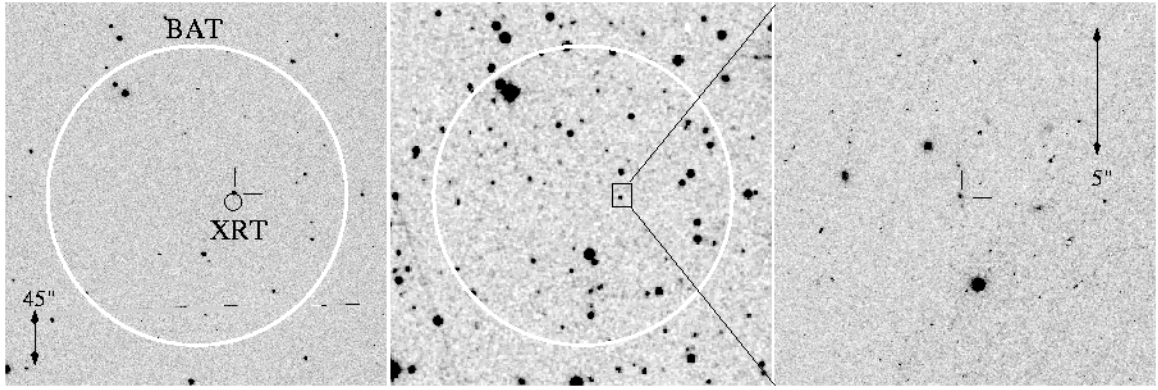}
\caption[$R$-band imaging of the field of GRB\,050820A]
        {$R$-band imaging of the field of GRB\,050820A.  \textit{Left}: P60
        $R_{\mathrm{C}}$-band
        discovery image of the afterglow of GRB\,050820A.  The BAT
        (2\arcmin-radius white circle) and XRT (7\arcsec-radius black circle)
        error circles are
        labeled.  The afterglow is identified with the two black tick marks.
        \textit{Center}: The Second-Generation Digitized Sky Survey image
        of the identical field.  The afterglow is not visible in this
        reference image.  \textit{Right}: \HST\ $F625W$ image of the afterglow
        (indicated again with the two black tick marks).  The two
        nearby objects complicating the ground-based photometry are visible
        (see \S\ref{sec:p60} for details).  All images
        are oriented with North up and East to the left.}
\label{fig:finder}
\end{figure*}

\subsubsection{\HST}
\label{sec:hst}
To better constrain the jet-break time,
as well as to investigate the properties of the host galaxy of GRB\,050820A,
we triggered our Cycle 14 \HST\ program (GO-10551; PI: Kulkarni).
Using the Wide-Field Camera (WFC) channel 
of the Advanced Camera for Surveys (ACS),
we imaged the field of GRB\,050820A on 29 September 2005 UT 
($t_{\mathrm{BAT}} \approx 37$ d)
in the $F625W$ ($r'$), $F775W$ ($i'$), and $F850LP$ ($z'$) filters
(see Figure \ref{fig:finder}).  A second
epoch to study the host galaxy is scheduled for July 2006, and these
results will be reported in a future work.  

The HST data were processed using the \texttt{multidrizzle} routine 
\citep{fh02} in the
\texttt{stsdas} IRAF package.  We used \texttt{pixfrac}$= 0.8$ and
\texttt{pixscale}$= 1.0$ for the drizzling procedure, resulting in a pixel
scale of $0.05$\arcsec\ / pixel.  The astrometry on these images was then 
tied to a P60 co-addition of all the $R_{\mathrm{C}}$-band data taken on 20
August 2005 (which is itself tied to the USNO-B1 astrometric grid).

The afterglow is well separated from any nearby objects in the field, and so we
have followed the recipe for aperture photometry from \citet{sjb+05}. 
As a note of caution, however, flux from an underlying host galaxy could
affect the results reported here.  Final values for the late time afterglow
flux will require image subtraction of any host galaxy contribution,
expected following our July observations.
$F625W$ and $F775W$ magnitudes were converted to the 
$R_{\mathrm{C}}$ and $I_{\mathrm{C}}$
band-passes using synthetic spectra from Table 22 of \citet{sjb+05}.
The results of our measurements are shown in 
Table~\ref{tab:optical}.

\subsection{Radio Observations}
\label{sec:radio}

\label{sec:vla}

In Table~\ref{tab:radio} we summarize our radio observations of 
GRB\,050820A, spanning 0.1--61 days after the explosion. All observations 
were conducted with the Very Large Array\footnote{The National Radio 
Astronomy Observatory is a facility of the National Science Foundation 
operated under cooperative agreement by Associated Universities, Inc.} 
in standard continuum mode with a bandwidth of 2 $\times$ 50\,MHz 
centered at 4.86, 8.46 or 22.5\,GHz.  The array was in the $C$ configuration,
with an angular resolution of 2.3\arcsec.
We used 3C48 (J0137$+$331) for flux 
calibration, while J2212+239 and J2225+213 were used to monitor phase. 
Data were reduced using standard packages within the Astronomical 
Image Processing System (AIPS).

\begin{deluxetable}{rrrc}
  \tabletypesize{\footnotesize}
  \tablecaption{Radio Observations of GRB\,050820A}
  \tablecolumns{4}
  \tablewidth{0pc}
  \tablehead{\colhead{Observation Date} & \colhead{$t_{\mathrm{BAT}}$} &
             \colhead{Frequency} & \colhead{Flux Density\tablenotemark{a}} \\
             \colhead{(2005 UT)} & \colhead{(days)} & \colhead{(GHz)} &
             \colhead{($\mu$Jy)}
            }
  \startdata
        Aug 20.39 & 0.116 & 4.86 & $< 102$ \\
        Aug 20.39 & 0.116 & 8.46 & $110 \pm 40$ \\
        Aug 20.39 & 0.116 & 22.5 & $< 186$ \\
        Aug 21.20 & 0.93 & 8.46 & $634 \pm 62$ \\
        Aug 22.42 & 2.15 & 4.86 & $256 \pm 78$ \\
        Aug 22.42 & 2.15 & 8.46 & $419 \pm 50$ \\
        Aug 22.42 & 2.15 & 22.5 & $< 216$ \\
        Aug 24.38 & 4.11 & 4.86 & $171 \pm 47$ \\
        Aug 24.38 & 4.11 & 8.46 & $74 \pm 36$ \\
        Aug 25.32 & 5.05 & 8.46 & $< 114$ \\
        Aug 26.40 & 6.13 & 8.46 & $< 120$ \\
        Aug 28.37 & 8.10 & 8.46 & $166 \pm 45$ \\
        Sep 1.33 & 12.06 & 8.46 & $89 \pm 39$ \\
        Sep 4.18 & 14.91 & 8.46 & $106 \pm 33$ \\
        Sep 15.20 & 25.93 & 8.46 & $76 \pm 30$ \\
        Oct 20.19 & 60.92 & 8.46 & $< 70$ \\
  \enddata
  \tablenotetext{a}{Errors quoted for detections are at the 1$\sigma$ level.
        Upper limits are reported as 2$\sigma$ RMS noise.}
\label{tab:radio}
\end{deluxetable}

\section{Analysis}
\label{sec:anal}

In this section we provide an analysis of the X-ray, optical,
and radio light curves and spectra of the afterglow of 
GRB\,050820A.  
We have divided the 
X-ray and optical light curves into segments 
(Phases 1--4) based on noticeable temporal
breaks.  We then investigated each segment independently, fitting the
light curve and spectra to power-law decay indices of the 
form: $F_{\nu}
\propto t^{-\alpha} \nu^{-\beta}$.  The lack of a bright afterglow makes
such analysis impossible in the radio.  

The X-ray and optical light curves, with temporal 
divisions marked as dashed vertical
lines, are shown in Figure \ref{fig:xolc}.  Phase 1 begins
with the BAT trigger and ends with the resumption of X-ray observations
at $t_{\mathrm{break,}1} \equiv 4785$ s.  The X-ray and optical data behave 
differently in Phase 1, resulting in unique subdivisions for the two 
band-passes.  However, Phase 1 is the only epoch to show such divergent 
behavior.

With the emergence of \Swift\ from the SAA at $t_{\mathrm{break,}1}$, both 
the X-ray and optical light curves 
exhibit a relatively shallow decline.  This characterizes Phase 2,
which ends when the X-ray decline steepens at $t_{\mathrm{break,}2} \equiv 8.7
\times 10^{4}$ s.  The decay slope in Phase 3 is steeper than in Phase 2
in both band-passes. 

Phase 3 extends out to the last X-ray detection at $t_{\mathrm{break,}3} \equiv 
1.7 \times 10^{6}$ s.  Between this time and the HST optical observations,
the optical decay must have significantly steepened.  This last epoch,
with only optical data, we
define as Phase 4.

The results of this power-law analysis are shown in Tables
\ref{tab:xrayab} and \ref{tab:optab}.  Each bandpass is discussed in further
detail below.

\begin{deluxetable*}{lrrccc}
  \tabletypesize{\footnotesize}
  \tablecaption{Optical Afterglow Spectral and Temporal Fits}
  \tablecolumns{6}
  \tablewidth{0pc}
  \tablehead{\colhead{Phase} & \colhead{$t^{\mathrm{start}}_{\mathrm{BAT}}$} &
             \colhead{$t^{\mathrm{stop}}_{\mathrm{BAT}}$} &
             \colhead{$\alpha$} &
             \colhead{$\beta$} &
             \colhead{$\chi_{\mathrm{r}}^{2}$ / d.o.f.} \\
             & (s) & (s) & & &
            }
  \startdata
        1a & $T_{\mathrm{BAT}}$ & 626 & $-0.35 \pm 0.02$ & \ldots &
                3.00 / 13 \\
        1b & 626 & $4.8 \times 10^{3}$ & $0.97 \pm 0.01$ & $0.57 \pm 0.06$
                & 1.53 / 14 \\
        2 & $4.8 \times 10^{3}$ & $8.7 \times 10^{4}$ & $0.78 \pm 0.01$ &
                $0.77 \pm 0.08$ & 5.7 / 31 \\
        3 & $8.7 \times 10^{4}$ & $1.7  \times 10^{6}$ &
                $0.99 \pm 0.06$ & $\ldots$ & 1.43 / 18 \\
        4 & $1.7 \times 10^{6}$ & $3.2 \times 10^{6}$ & $\ge 2.1$ &
                $\ldots$ & $\ldots$ \\
  \enddata
  \tablecomments{We have fit the optical data to a power-law model of the
        form $F_{\nu} \propto t^{-\alpha} \nu^{-\beta}$ where possible.
        In some Phases (1a and 3), we have limited spectral coverage
        and could not meaningfully constrain $\beta$.  Thus we have only
        fit for the temporal decay index $\alpha$.  In Phase 4, we can only
        place an upper limit on the decay slope.  All errors quoted are
        90\% confidence limits.}
\label{tab:optab}
\end{deluxetable*}
 

\subsection{X-ray Light Curve and Spectrum}
\label{sec:xraylc}

The X-ray light curve of GRB\,050820A is shown in Figure \ref{fig:xolc}
(top panel).
In Phase 1, we see two distinct behaviors: initially the X-ray light curve
falls rapidly with a decay slope $\alpha_{1\mathrm{a,x}} = 2.2 \pm 0.3$.
This continues until $t_{BAT} = 217$ s, and we define this as Phase
1a$_{\mathrm{x}}$. 
The X-ray emission then rises rapidly ($217 < t_{\mathrm{BAT}}
< 257$ s; Phase 1b$_{\mathrm{x}}$), after which \Swift\ enters the SAA.
It is clear from the correlation between the $\gamma$-ray and X-ray emission
in this epoch that the two are related (see Fig.~\ref{fig:early_all}c).
A full study of the properties of Phase 1b$_{\mathrm{x}}$ is left to
\S\ref{sec:early}.  

Following \Swift's emergence from the SAA, 
the light curve in Phase 2 shows evidence
for a shallower epoch of decline.  With the large gap in coverage, we cannot
constrain when this transition occurs.  We therefore define 
$t_{\mathrm{break,}1}$ to coincide with the resumption of XRT observations
at $t_{\mathrm{BAT}} = 4785$ s.  
A similar break in the optical light curve is also seen near this time 
(\S\ref{sec:optlc}).  

The X-ray data after $t_{\mathrm{break,}1}$ are not well fit by a single
power-law decay ($\chi^{2}_{\mathrm{r}} = 3.7$; 31 d.o.f.), 
due mostly to a steepening
of the decay at $t_{\mathrm{BAT}} \approx 10^{5}$ s.  
Fitting a broken power-law
model to this data, we find an acceptable fit with $t_{\mathrm{break,}2} =
(8.7 \pm 2.4) \times 10^4$ s ($\chi^{2}_{\mathrm{r}} = 1.19$; 25 d.o.f.). 
The resulting
decay index before the break (Phase 2) is 
$\alpha_{2,\mathrm{x}} = 0.93 \pm 0.03$.  For Phase 3, we find
$\alpha_{3,\mathrm{x}} = 1.25 \pm 0.07$. 

The X-ray spectral index in Phase 1a$_{\mathrm{x}}$ is relatively steep:
$\beta_{1\mathrm{a,x}} = 0.90 \pm 0.09$.  The spectrum hardens
significantly in Phase 1b$_{\mathrm{x}}$ ($\beta_{1\mathrm{b,x}} = 
-0.10 \pm 0.03$), further justifying our decision to split Phase 1
into two separate X-ray segments.  In Phase 2, the spectrum softens again, to
$\beta_{2\mathrm{,x}} = 1.20 \pm 0.04$.  There are too few 
X-ray counts in Phase 3 to meaningfully constrain the spectrum.

The results of our analysis of the X-ray data set are summarized in
Table \ref{tab:xrayab}.

\subsection{Optical Light Curve and Spectrum}
\label{sec:optlc}

\begin{figure*}
\epsscale{1.0}
\plottwo{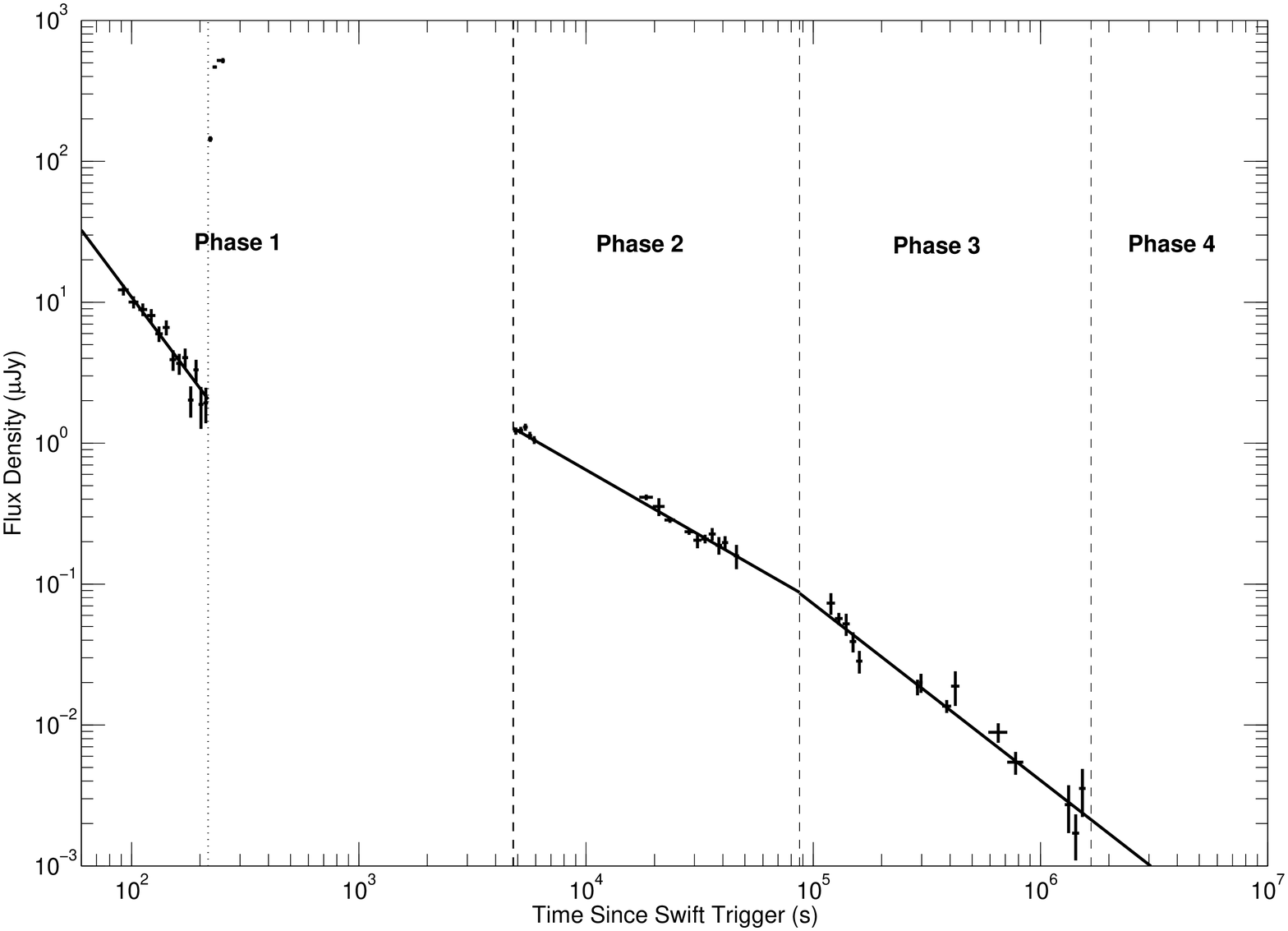}{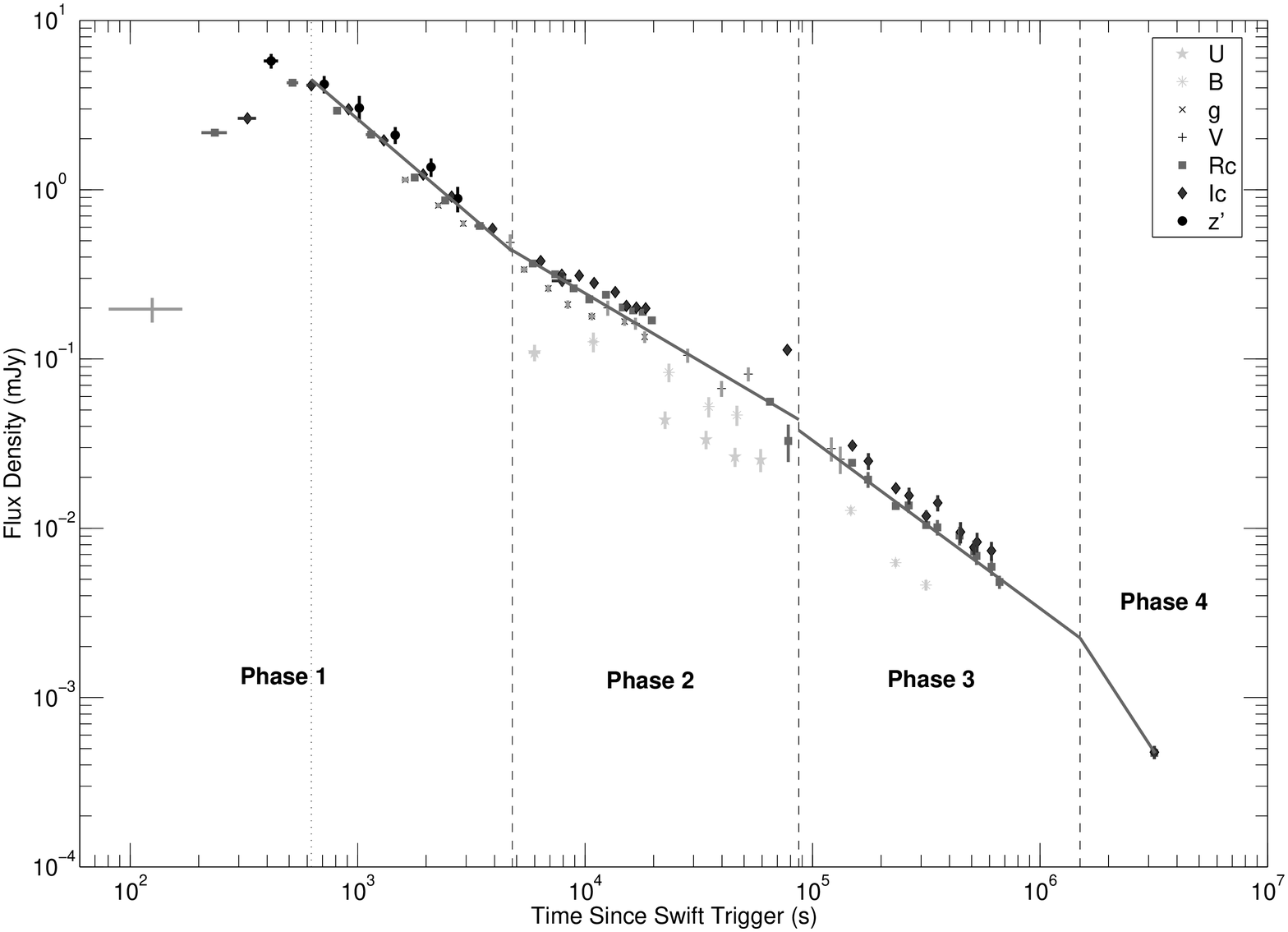}
\caption{X-ray (top) and optical (bottom) light curves of GRB\,050820A.
        Both band-passes have been divided into four segments (Phases 1--4),
        each shown with a vertical dashed line.  The unique subdivision of
        Phase 1 is shown as a dotted line in both plots.
        \textit{Top:} 2--10 keV X-ray fluxes were converted to flux densities
        at 5 keV using the average spectral slope for each Phase.  It was
        assumed the spectrum remained constant from Phase 2 to Phase 3.
        The best-fit X-ray temporal decay is shown with a solid line.
        \textit{Bottom:} Magnitudes were converted to flux densities using
        zero-points from \citet{fsi95}.  The best-fit $R_{\mathrm{C}}$-band
        temporal decay is shown with a solid line.
        For full details of our analysis, see \S\ref{sec:anal} and Tables
        \ref{tab:xrayab} and \ref{tab:optab}.}
\label{fig:xolc}
\end{figure*}

The optical light curve from GRB\,050820A is shown in Figure
\ref{fig:xolc} (bottom panel).  
Phases 1 ($0 < t_{\mathrm{BAT}} < 4785$), 2 ($4785 < t_{\mathrm{BAT}} <
8.7 \times 10^{4}$), and 3 ($8.7 \times 10^{4} < t_{\mathrm{BAT}} <
1.7 \times 10^{6}$) of the optical light curve have already been defined
in terms of the X-ray decay.  However, unlike the X-ray, the earliest
optical observations indicate the afterglow was getting brighter
with time 
\citep{GCN.3834, GCN.3836}.  This rise continues until a peak at 
$t_{\mathrm{BAT}} \approx 600$ s, marking the end of 
Phase 1a$_{\mathrm{opt}}$.  After the peak, the optical light curve
in all four P60 filters decays steadily with $\alpha_{1\mathrm{b,opt}} = 
0.97 \pm 0.01$ until $t_{\mathrm{break,1}}$.  We note that for this
and all subsequent phases, we have constrained $\alpha$ to be
identical in all optical filters.  A more thorough discussion
of Phase 1 is left to \S\ref{sec:early}.

Much like the X-ray, the optical decay in Phases 2 and 3 is poorly fit
by a single power-law ($\chi^{2}_{\mathrm{r}} = 13.7$; 49 d.o.f.).  
In Phase 2, the optical decay noticeably flattens ($\alpha_{2,\mathrm{opt}} =
0.78 \pm 0.01$).  A much higher degree
of variability is seen in the different filters, resulting in a 
poor fit statistic.  After $t_{\mathrm{break,2}}$
the decay in Phase 3 is again steeper and more uniform, with 
$\alpha_{3,\mathrm{opt}} = 0.99 \pm 0.06$.  

It is clear that 
if we extrapolate the decay from Phase 3 out to the HST observations, 
the late-time flux is greatly overestimated.  We conclude therefore a break
has occurred in the light curve sometime after $t_{\mathrm{break,3}}$, and
thus we define Phase 4 to span $1.7 \times 10^{6} < t_{\mathrm{BAT}} <
3.2 \times 10^{6}$.  We estimate the temporal decay in Phase 4 to be
$\alpha_{4,\mathrm{opt}} \geq 2.1$.

Due to the limited spectral coverage of our observations,
we are unable to provide meaningful constraints on the spectral index 
$\beta$ in Phases 1a$_{\mathrm{opt}}$, 3, and 4.  For the remaining epochs,
we have excluded the $U$- and $B$-band data from our spectral 
fits, as these are 
expected to lie below the Ly-$\alpha$ absorption edge at this redshift.  
We attempted to solve for the host galaxy reddening ($A_{V}$(host))
using extinction laws for the Milky Way, Large and Small Magellanic Clouds 
from \citet{p92}.  In all cases, we find a host extinction consistent
with zero.  While this result is inconsistent with the column density 
derived from high-resolution spectroscopy \citep{GCN.3860}, it is in agreement
with the low host extinction values seen in almost all well-sampled 
pre-\Swift\ afterglows \citep{kkz06}.  

Ignoring host reddening, we find 
$\beta_{1\mathrm{b,opt}} = 0.57 \pm 0.06$ and 
$\beta_{2\mathrm{,opt}} = 0.77 \pm 0.08$.  
While the optical spectrum appears to have steepened in Phase 3, the poor fit
quality of this phase precludes any firm conclusions from being drawn.

The results of our analysis of the optical data set are shown in 
Table \ref{tab:optab}. 

\subsection{Radio Light Curve}
\label{sec:radiolc}

\begin{figure}
\epsscale{1.0}
\plotone{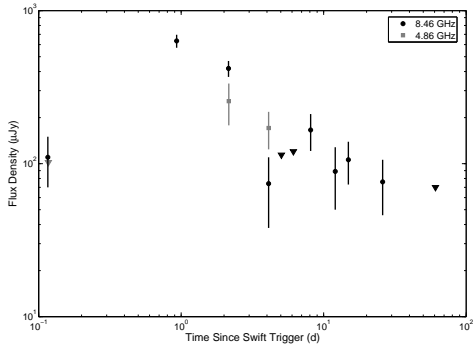}
\caption{Radio Afterglow of GRB\,050820a.  The early rise in the radio light
        curve at $t_{\mathrm{BAT}} \approx 1$ d is most easily understood
        as a reverse shock caused by late-time energy injection, as seen
        in both the optical and X-ray light curves.
        The most striking feature of the radio light curve,
        however, is the lack of a bright radio afterglow at late times
        (see \S\ref{sec:cbm}).}
\label{fig:radio}
\end{figure}

The radio emission rises to a peak sometime around 1 d
after the burst (see Figure ~\ref{fig:radio} and Table~\ref{tab:radio}).  
The radio spectrum at early times is quite chaotic,
transitioning through a peak around 8.5 GHz at $t_{\mathrm{BAT}} = 2.15$ d
to optically thin 4.11 d after the burst.  We note that some of the
variation at early times may be due to inter-stellar 
scintillation \citep{g97} as has been
in seen in many other radio afterglows (see e.g.~\citealt{fkn+97}).

The late-time ($t > 7$ d) radio data show no sign of any afterglow brighter 
than 200 $\mu$Jy.
This is in marked contrast to the bright optical and X-ray afterglows, and
is one of the most difficult aspects of the afterglow to account for
(\S\ref{sec:cbm}).

\section{Discussion}
\label{sec:disc}

In this section, we use the results from our previous analysis to try
and explain the broadband emission from GRB\,050820A in the context of
the standard fireball model (see e.g.~\citealt{p05} for a review).
In this model, highly-relativistic
ejecta are emitted by a central engine as shells with varying Lorentz 
factors.  Shock fronts formed between these shells produce the high-energy
prompt emission; these collisions are known as internal shocks.  Emission
from internal shocks is highly non-thermal and results in  
an (empirically determined) power-law spectrum with a cut-off exponential:
$\mathrm{d}N / \mathrm{d}E \propto E^{\alpha} e^{-E / E_{0}}$
\citep{bmf+93}.  As the relativistic 
shells expand and slow down, they eventually encounter the circum-burst
medium.  Again a collision-less shock front forms, accelerating electrons 
to a power-law distribution of Lorentz factors with exponent $p$ and 
minimum Lorentz factor $\gamma_{\mathrm{m}}$.  It is assumed that a constant 
fraction of the total energy density is partitioned to the 
electrons ($\epsilon_{e}$) and the magnetic field ($\epsilon_{B}$).
These accelerated electrons 
then emit synchrotron radiation, powering the long-lived X-ray, optical,
and radio afterglow (the forward, external shock).  Additional emission can
be generated from shock heating of the ejecta (the reverse shock), leading
to a rapidly-decaying flare in the optical and radio light curve \citep{sp99}.

The observed afterglow spectrum depends on the relative ordering of 
the three critical frequencies: $\nu_{\mathrm{a}}$, the frequency
where self-absorption becomes important, $\nu_{\mathrm{m}}$, the characteristic 
frequency of the emission, and $\nu_{\mathrm{c}}$, the frequency above
which electrons are able to cool efficiently through radiation.  Typical
afterglow observations occur when $\nu_{\mathrm{a}} < \nu_{\mathrm{m}} <
\nu_{\mathrm{c}}$ (i.e.~slow cooling), resulting in the following spectral
indices \citep{spn98}:
\begin{eqnarray}
\label{eqn:nu1a}
F_{\nu} \propto & \nu^{2} & ;\: \nu < \nu_{\mathrm{a}} \\
\label{eqn:nu1b}
		& \nu^{1/3} & ;\:\nu_{\mathrm{a}} < \nu < \nu_{\mathrm{m}} \\
\label{eqn:nu1c}
		& \nu^{-(p-1)/2} & ;\:\nu_{\mathrm{m}} < \nu < 
				    \nu_{\mathrm{c}} \\
\label{eqn:nu1d}
		& \nu^{-p/2} & ;\: \nu > \nu_{\mathrm{c}}
\end{eqnarray}

The light curve produced by such emission depends on the 
radial profile of the circum-burst medium into which the shock is expanding.
The simplest circum-burst medium to consider is one in which 
the density is constant ($\rho \propto r^{0}$).  This scenario is also
referred to as an inter-stellar medium (ISM).  In this case, the flux
density will scale as \citep{spn98}:
\begin{eqnarray}
\label{eqn:t1a}
F_{\nu} \propto & t^{1/2} & ;\: \nu < \nu_{\mathrm{a}} \\
\label{eqn:t1b}
		& t^{1/2} & ;\: \nu_{\mathrm{a}} < \nu  < \nu_{\mathrm{m}} \\ 
\label{eqn:t1c}
		& t^{3(1-p)/4} & ;\: \nu_{\mathrm{m}} < \nu < 
				 \nu_{\mathrm{c}} \\
\label{eqn:t1d}
		& t^{(2-3p)/4} & ;\: \nu > \nu_{\mathrm{c}}
\end{eqnarray}
Alternatively, if we eliminate $p$ from the above equations, we find 
a characteristic relation between $\alpha$ and $\beta$ in each spectral 
regime known as a ``closure relation'' \citep{pbr+02}:
\begin{eqnarray}
\label{eqn:c1a}
\alpha = & \frac{\beta}{4} & ;\: \nu < \nu_{\mathrm{a}} \\
\label{eqn:c1b}
	 & \frac{3 \beta}{2} & ;\: \nu_{\mathrm{a}} < \nu < \nu_{\mathrm{m}} \\
\label{eqn:c1c}
	 & \frac{3 \beta}{2} & ;\: \nu_{\mathrm{m}} < \nu < \nu_{\mathrm{c}} \\
\label{eqn:c1d}
	 & \frac{3 \beta - 1}{2} & ;\: \nu > \nu_{\mathrm{c}}
\end{eqnarray}

The long-soft class of GRBs, however, 
is thought to arise from the deaths of massive
stars, as they collapse to form black-holes: the so-called ``collapsar''
model \citep{w93}.  In the late stages of evolution,
massive Wolf-Rayet stars are stripped of their outer envelopes in a wind,
leaving behind a signature $\rho \propto r^{-2}$ density profile that
should be discernible in the afterglow light curve.  The analogous temporal
decay indices for a wind-like medium are \citep{cl00}:
\begin{eqnarray}
\label{eqn:t2a}
F_{\nu} \propto & t^{1} & ;\: \nu < \nu_{\mathrm{a}} \\
\label{eqn:t2b}
                & t^{0} & ;\: \nu_{\mathrm{a}} < \nu < \nu_{\mathrm{m}} \\
\label{eqn:t2c}
                & t^{(1-3p)/4} & ;\: \nu_{\mathrm{m}} < \nu < 
				 \nu_{\mathrm{c}} \\
\label{eqn:t2d}
                & t^{(2-3p)/4} & ;\: \nu > \nu_{\mathrm{c}}
\end{eqnarray}
The derived closure relations are:
\begin{eqnarray}
\label{eqn:c2a}
\alpha = & \frac{\beta}{2} & ;\: \nu < \nu_{\mathrm{a}} \\
\label{eqn:c2b}
         & \frac{3 \beta + 1}{2} & ;\: \nu_{\mathrm{a}} < \nu < 
				   \nu_{\mathrm{m}} \\
\label{eqn:c2c}
         & \frac{3 \beta + 1}{2} & ;\: \nu_{\mathrm{m}} < \nu < 
	                           \nu_{\mathrm{c}} \\
\label{eqn:c2d}
         & \frac{3 \beta - 1}{2} & ;\: \nu > \nu_{\mathrm{c}}
\end{eqnarray}

The above temporal decay indices and closure relations
(Equations \ref{eqn:t1a}-\ref{eqn:c2d}) are only valid
for spherically symmetric emission.  GRBs, however, are thought to be 
beamed events.  At early times, observers only notice emission from a narrow
cone (opening angle $\theta \sim \Gamma^{-1}$, where $\Gamma$ is the Lorentz
factor of the expanding shock) due to relativistic beaming 
(see e.g.~\citealt{rl}).   
As the shock slows,
however, lateral spreading of the jet becomes important, and the observer
eventually notices ``missing'' emission from wider angles \citep{r99,sph99}.
This hydrodynamic transition manifests itself as an achromatic steepening
in the afterglow light curve, with an expected post-jet-break decay 
proportional to $t^{-p}$.  

With the above formulation in hand, we now wish to understand the 
physical implications of our previous analysis.  


\subsection{Early $\gamma$-ray Emission}
\label{sec:pre}
The most striking feature of the $\gamma$-ray light curve of GRB\,050820A
is the large gap between the initial pulse that triggered the \Swift\ BAT
(Peak A in Fig.~\ref{fig:early_all}b)
and the bulk of the high-energy emission ($t_{\mathrm{BAT}} > 200$ s).  
The natural
question arises as to whether this ``precursor''\footnote{Here we define
a precursor as an event that is well separated from and contains only a small
fraction of the total high-energy emission.  Unlike some other authors, our 
definition is independent of the mechanism behind the emission.  Peak
$A$ in GRB\,050820A is then clearly a precursor.} results from the same
physical mechanism as the bulk of the high energy emission.  Many 
models predict a high-energy component distinct from the prompt GRB 
at early times.  Possible mechanisms include the
transition from an optically thick to an optically thin environment in the
fireball itself \citep{p86,lu00,mr00,lb03}, 
or the interaction of a jet with a progenitor, 
presumably a collapsing Wolf-Rayet star \citep{rml02,wm03}.  

We can securely rule out both of these models for the precursor of 
GRB\,050820A, as both predict a thermal spectrum.  Fitting a thermal model
to the BAT precursor spectrum results in a fit statistic of 
$\chi^{2}_{\mathrm{r}} = 
3.4$ (75 d.o.f.), while a non-thermal power-law model
provides an excellent fit ($\chi^{2}_{\mathrm{r}} = 1.07$ for 75 d.o.f.;  
$\Gamma = 1.74 \pm 0.08$).

While the precursor may be non-thermal, it is noticeably softer than the 
majority of the remaining prompt emission (see Figs.~\ref{fig:kwhr}d and
\ref{fig:kwhr}e).  
A search for precursors in a sample of long, bright \BATSE\ bursts revealed
such a soft, non-thermal component in a sizable fraction (20--25\%)
of these events \citep{l05}.  Furthermore, two of the longest, brightest 
\Swift\ bursts observed to date, GRB\,041219 \citep{vww+05,mhm+06} 
and GRB\,060124
\citep{rcc+06} show a faint, soft precursor followed by a large time lag
(570 s in the case of GRB\,060124).  

These soft precursors
are inconsistent with the main prompt
emission in most GRBs,
which exhibits a hard-to-soft evolution in the $\gamma$-ray
spectrum \citep{fbm+95,fac+00}.  The $\gamma$-ray light curve of GRB\,050820A
conforms to this trend only if we ignore the 
precursor.  Furthermore, it is difficult to conceive
of a scenario by which internal shocks can generate such long periods
of quiescence in a sustained outflow.  The large time lag, soft nature,
and repeated occurrence of these precursors hint that they are in fact
due to a different emission mechanism than the internal dissipation thought
to power the bulk of the high-energy emission.
However, we cannot state this conclusively, as would be the case
if the precursors were thermal.

If we assume 
a different emission mechanism, the prompt emission 
did not begin until 222 s after the \Swift\ trigger.  
This seemingly small discrepancy in 
defining $T_{0}$ affects the calculated 
temporal decay indices,
particularly during the early afterglow (\S\ref{sec:cbm}).  
For all temporal decay indices 
calculated in this work, we consider $T_{0}$ to coincide with the 
beginning of the bulk of the
high-energy emission (i.e.~06:38:35 UT on 20 August 2005).

Finally, we consider the early X-ray emission.  The temporal decay slope at
early times ($\alpha_{1\mathrm{a,X}} = 2.2$) is too steep to be attributed
to a standard forward shock afterglow.  The most popular
explanation for the rapid decline of early X-ray emission in \Swift\ GRBs is
``high-latitude emission'':  prompt emission from large angles ($\theta >
\Gamma^{-1}$) that, due to relativistic beaming effects, reaches the observer
at late times ($\Delta t \sim (1 + z) R \theta^{2} / 2 c$, \citealt{kp00a}).
However, this results in a well-defined relationship between the spectral
and temporal indices ($\alpha = \beta + 2$) which is inconsistent with the
observed values for GRB\,050820A.  

\citet{zfd+06} discuss possible mechanisms
that could cause the early-time decay slope to deviate from this behavior.
The most realistic possibility is if the X-rays were below the cooling
frequency at this very early epoch.  Then the closure relation would
take the form $\alpha \approx 1 + 3 \beta / 2 = 2.4$ \citep{sp99b}, in good
agreement with the observed value.


\begin{figure}
\epsscale{1.0}
\plotone{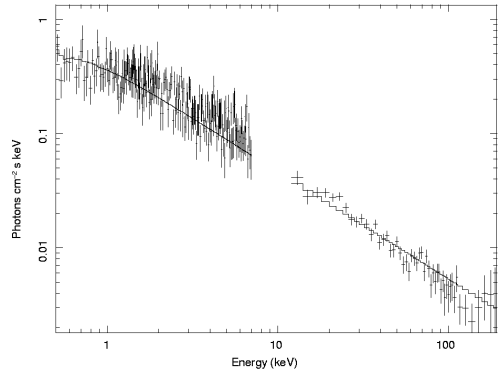}
\caption{Joint BAT/XRT Spectrum of the Main Pulse of Prompt Emission
        ($217 < t_{\mathrm{BAT}} < 257$ s).  While there is no region of
        direct overlap, the XRT data is clearly the low-energy tail of the
        prompt BAT emission, forming one continuous spectrum.  The best
        fit spectrum ($\Gamma = 0.94$) is shown as a solid line.  Both the
        BAT and XRT data have been binned for plotting purposes.}
\label{fig:batxrt}
\end{figure}

\begin{deluxetable*}{lrrccccrcc}
  \tabletypesize{\footnotesize}
  \tablecaption{Joint $\gamma$-ray / Optical Early-Time Data}
  \tablecolumns{9}
  \tablewidth{0pc}
  \tablehead{\colhead{Interval ID} &
             \colhead{$t_{\mathrm{BAT}}^{\mathrm{start}}$} &
             \colhead{Duration} &
             \colhead{$\gamma$-Ray Flux\tablenotemark{a}} &
             \colhead{$\alpha$\tablenotemark{b}}
             & \colhead{$E_{\mathrm{p}}$\tablenotemark{b}} &
             \colhead{$\chi_{\mathrm{r}}^{2}$ / d.o.f.} &
             \colhead{Optical Filter}
             & \colhead{Optical Flux Density\tablenotemark{a}} &
             \colhead{$C_{\mathrm{O}
              \gamma}$\tablenotemark{c}} \\
             & (s) & (s) & ($10^{-8}$ erg cm$^{-2}$ s$^{-1}$) & &
             (keV) &  &  & (mJy)
            }
  \startdata
        1 & 80.949 & 85.376 & $< 1.3$ & $\ldots$ & $\ldots$ & $\ldots$ &
                $V$ & $0.20 \pm 0.03$ & $< 10.2$ \\
        2 & 257.839 & 8.448 & $96.2^{+13.9}_{-0.80}$ & $1.26 \pm 0.14$ &
                $510^{+211}_{-120}$ & 0.83 / 62 & $R_{\mathrm{C}}$ & $2.17 \pm
                0.08$ & 12.5 \\
        3 & 297.775 & 59.904 & $< 3.2$ & $\ldots$ & $\ldots$ & $\ldots$ &
                $I_{\mathrm{C}}$ & $2.64 \pm 0.04$ & $< 8.8$ \\
        4 & 389.167 & 57.344 & $19.06^{+0.03}_{-6.61}$ & $1.13^{+0.24}_{-0.29}$
                & $269^{+107}_{-59}$ & 0.78 / 62 & $z'$ & $5.77 \pm
                0.57$ & 10.0 \\
        5 & 487.471 & 57.344 & $9.35^{+0.53}_{-2.38}$ & $1.96 \pm 0.18$
                & $\ldots$ & 1.01 / 58 & $R_{\mathrm{C}}$ & $4.28 \pm
                0.08$ & 9.2 \\
        6 & 602.159 & 49.152 & $< 2.3$ & $\ldots$ & $\ldots$ & $\ldots$ &
                $I_{\mathrm{C}}$ & $4.14 \pm 0.05$ & $< 7.9$ \\
        7 & 684.079 & 57.344 & $< 4.1$ & $\ldots$ & $\ldots$ & $\ldots$ &
                $z'$ & $4.20 \pm 0.49$ & $< 8.7$ \\
  \tablenotetext{a}{Errors quoted are at the 1$\sigma$ level.}
  \tablenotetext{b}{Spectral fits of the form $\mathrm{d}N / \mathrm{d}E
        \propto E^{-\alpha}
        \exp^{-(2-\alpha) E / E_{\mathrm{p}}}$ were performed for the case
        of Intervals 2 and 4.  For interval 5, the highest-energy
        data were not of
        sufficient quality to estimate $E_{\mathrm{p}}$.  Instead a power-law
        fit ($\mathrm{d}N / \mathrm{d}E \propto E^{-\alpha}$) was used.
        Errors quoted are 90\% confidence limits.}
  \tablenotetext{c}{$\gamma$-ray-to-optical color index: $C_{\mathrm{O}\gamma}
        \equiv -2.5 \log (F(\mathrm{opt}) / F(\gamma))$.}
  \enddata
\label{tab:early}
\end{deluxetable*}

\subsection{Contemporaneous X-ray and Optical Emission}
\label{sec:early}
Given the long duration and bright fluence, GRB\,050820A provides a rare
opportunity to study contemporaneous emission in the optical, X-ray, and 
$\gamma$-ray band-passes.  In Figure \ref{fig:early_all}, we show the early
time ($t_{\mathrm{BAT}} \lesssim 800$ s)
emission in X-rays (c) and optical (d) overlaid onto the high-energy
light curve of GRB\,050820A.

A look at the X-ray data in Figure \ref{fig:early_all}c shows a strong
correlation between the X-ray and $\gamma$-ray light curves.  The 
X-ray light curve, previously in the midst of a decline, abruptly
jumps in sync with the high energy emission at $t_{BAT} \approx 222$ s (Phase
1b$_{\mathrm{X}}$).
In addition to temporal similarities, the X-ray photon index at this epoch
($\Gamma_{\mathrm{XRT}} =
0.90 \pm 0.03$) is much harder than at any other epoch in the X-ray light 
curve, and similar to that derived from the BAT
($\Gamma_{\mathrm{BAT}} = 1.07 \pm 0.06$).  
Thus motivated, we have performed a joint fit of the BAT and XRT spectra
at this epoch.  Unfortunately \KW\ had yet to trigger, and so no high-energy
multi-channel spectra are available from that instrument.
We find both band-passes well-fit by a single power-law
with index $\Gamma = 0.94 \pm 0.03$ ($\chi_{\mathrm{r}}^{2} = 1.3$; 
391 d.o.f.).  The resulting un-folded spectrum is shown in Figure 
\ref{fig:batxrt}.  We conclude
the X-rays in Phase 1b$_{\mathrm{X}}$ are generated by the same mechanism
as the prompt emission. 

It is clear from 
Figure \ref{fig:early_all}d that, unlike in the X-ray band, there is no
strong correlation between optical and $\gamma$-ray flux from GRB\,050820A.
Radical spectral evolution would be required in the optical to explain
both band-passes as arising from the same emission mechanism.  We consider
this scenario highly unlikely and conclude that, at the very least, the 
dominant contribution to the optical emission in Phase 1 has a different
origin than the prompt emission.

We next consider if our optical observations in Phase 1 can be explained
solely in terms of the standard afterglow formulation.  
We have attempted to fit both a simple
broken power-law \citep{spn98} as well as an analytic solution for the
flux density near the optical peak \citep{gs02} to our $R_{\mathrm{C}}$-,
$I_{\mathrm{C}}$-, and $z'$-band early-time
data.  The resulting fit quality is quite poor ($\chi_{\mathrm{r}}^{2} = 3.0$;
14 d.o.f.) with
the dominant contribution coming from the data in Phase 1a$_{\mathrm{opt}}$
(i.e.~before the peak).
In spite of the poor fit, we have included
the results for Phase 1a$_{\mathrm{opt}}$ in Table \ref{tab:optab} for
reference.

This result is not unexpected, as \citet{vww+06}
have shown that contemporaneous optical imaging of GRB\,050820A with the 
\raptor\ telescope can be well-fit as the sum of two independent components:
one representing the forward shock and another proportional to the high-energy
prompt emission.   We attempted an analogous fit with the P60 and \KW\ data
set.  While a better fit statistic ensues, we still do not find an 
acceptable result ($\chi^{2}_{\mathrm{r}} = 2.2$; 14 d.o.f.).  
We conclude that the
relatively sparse time sampling of our observations, 
coupled with the frequent filter changes,
make it impossible to verify this result.

Independent of any correlation between the prompt optical and $\gamma$-ray 
emission, we note that the decay after $t_{\mathrm{BAT}} = 600$ s is dominated
by the forward shock.  Unlike the bright, early-time emission seen from 
GRB\,990123 \citep{abb+99}, we see no evidence for rapidly-decaying 
($\alpha_{\mathrm{RS}} = (27p +7) / 35 \approx 2$; \citealt{k00})
reverse-shock emission from an optical flare.

\begin{figure*}
\epsscale{0.35}
\plotone{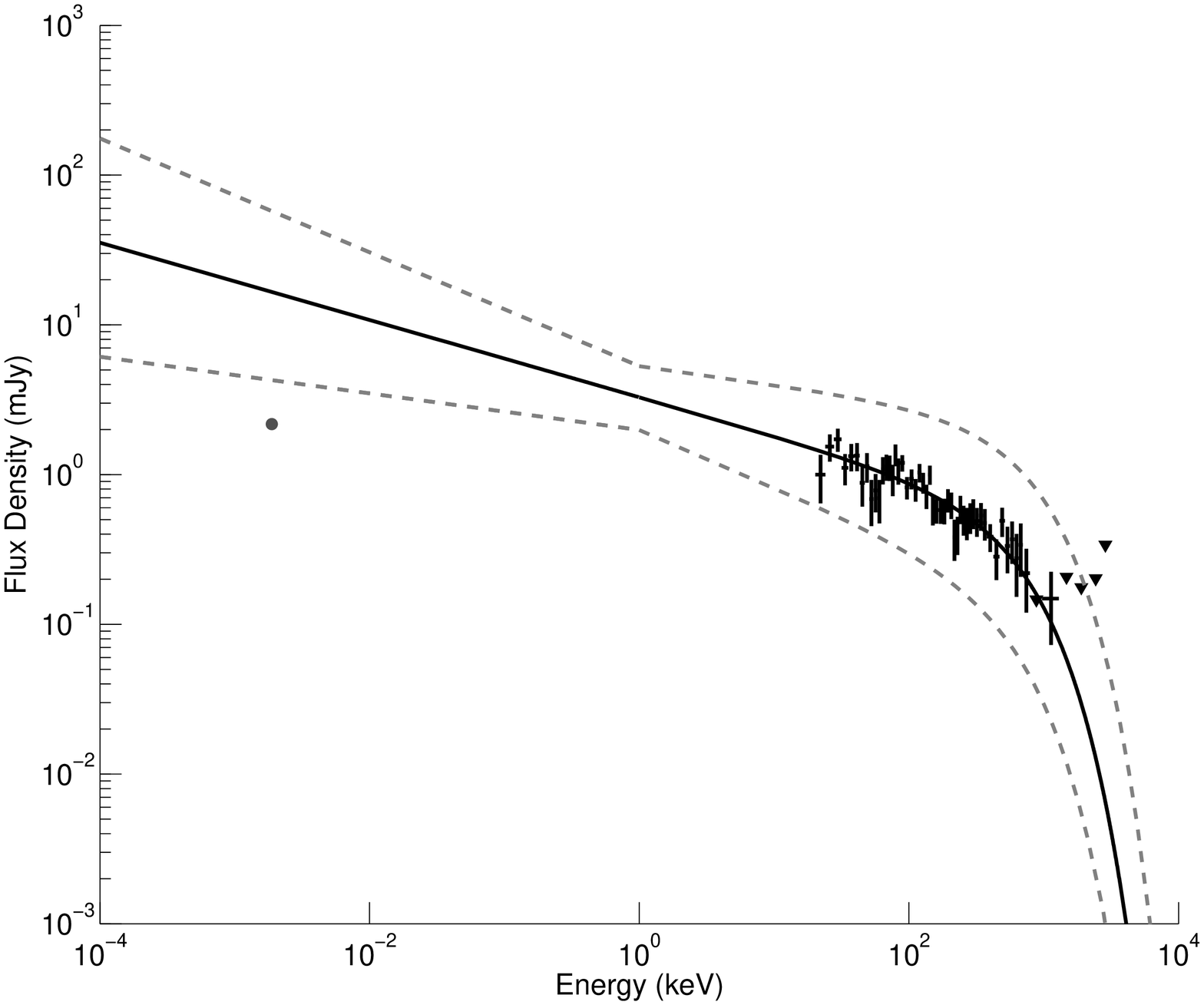}
\plotone{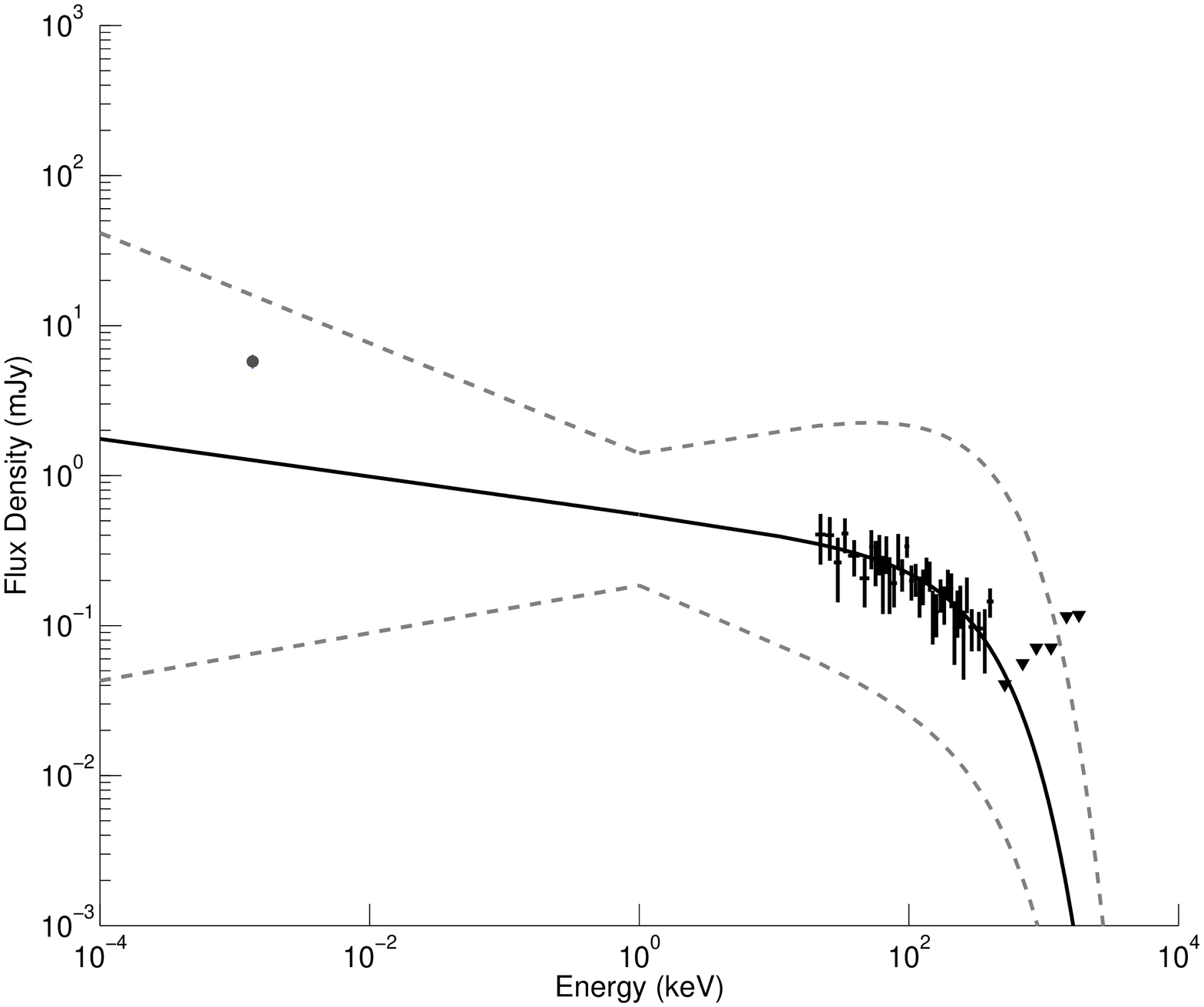}
\plotone{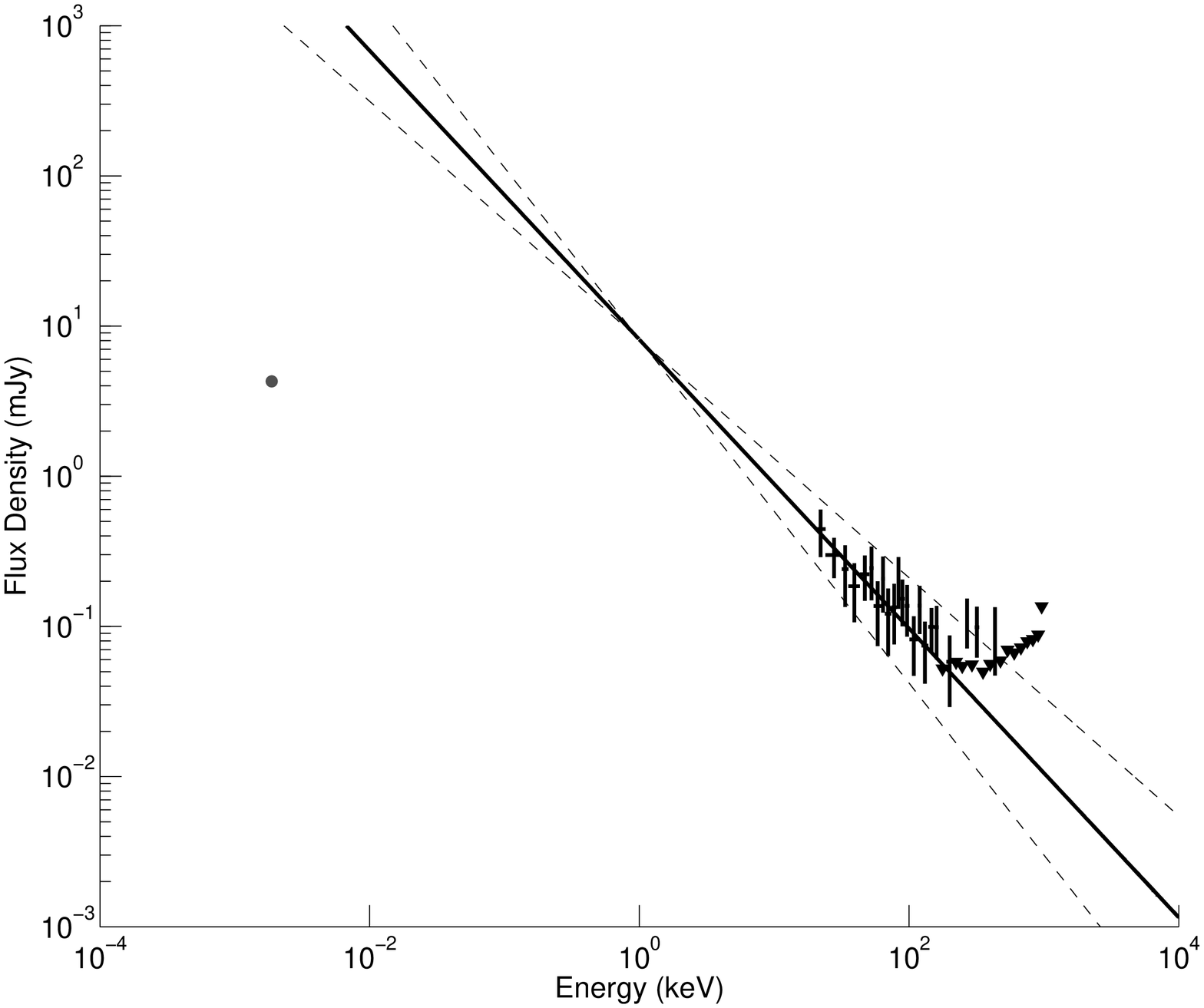}
\caption{Early-time Broadband Spectral Energy Distribution of GRB\,050820a.
 \KW\ spectral data for the given intervals (see Table \ref{tab:early} for
 definitions of the intervals) are shown (crosses) alongside the corresponding
 optical observations (circles).  2-sigma upper-limits in the high-energy
 spectra are plotted as triangles.  The best-fit model to the \KW\ spectrum
 is shown as a solid line, and the dashed lines show the 90\% confidence
 intervals for the spectral fits.  The ratio of optical to $\gamma$-ray
 flux varies significantly between the three intervals.
 \textit{Left}: Interval 2.  This interval covers only a small fraction of the
 time of the corresponding P60 image because \KW\ only triggered on GRB\,050820a (and hence began collecting multi-channel spectra) in the middle of this
 image.  The optical $R_{C}$-band data point lies below the extrapolation
 of the $\gamma$-ray spectrum.
 \textit{Center}: Interval 4.  Here the \KW\ and P60
 intervals are nearly simultaneous.  Unlike the other intervals,
 the P60 $z'$-band point lies above the predicted value and
 within the 90\% confidence interval of the
 extrapolation of the high-energy spectrum.
 \textit{Right}: Interval 5.
 Here the high-energy extrapolation greatly overestimates the optical
 flux.
 However, the $\gamma$-ray
 flux in this interval is quite low, and in particular the low number
 of high-energy photons makes it difficult to constrain a cut-off power-law
 spectrum.  In fact this interval was best fit with a simple power-law
 spectrum.}
 \label{fig:earlysed}
 \end{figure*}

Finally, using the combination of optical and $\gamma$-ray data, we consider
the broadband spectral energy distribution (SED) 
of GRB\,050820A at early times.  For each of the 
contemporaneous optical observations, we have extracted fluxes and 
(where possible) spectra from the corresponding \KW\ observations (Figure
\ref{fig:earlysed} and Table \ref{tab:early}).  
Following the method of \citet{vww+05}, we have calculated the 
$\gamma$-ray-to-optical color index, $C_{\mathrm{O}\gamma} \equiv
-2.5 \log (F(\mathrm{opt}) / F(\gamma))$, or lower limits, for each
interval. 
The ratio varies significantly over the course of
our observations.  The value of $C_{\mathrm{O}\gamma} = 12.5$ in Interval
2 is consistent with that seen from GRB\,041219a, while later intervals
are even brighter in the optical.  In fact, the optical-to-$\gamma$-ray flux
ratio in Interval 5 is over 240 times larger than that observed for
GRB\,050401 \citep{ryk+05}.  Evidently a large diversity exists in the 
broadband SEDs of GRBs at early times. 

\subsection{Late-time Energy Injection}
\label{sec:inject}
The majority of XRT light curves observed to date have exhibited a period
of shallow decline ($0.2 \lesssim \alpha \lesssim 0.8$) that is inconsistent
with the standard afterglow formulation \citep{nkg+06}.  Two models have been
invoked to explain this phase, both of which involve injecting energy 
into the forward shock at late times ($t >> t_{\mathrm{GRB}}$; 
see e.g.~\citealt{zfd+06}
for a review).  In the first, the central engine is active for long
time periods, $t >> t_{\mathrm{GRB}}$.  The late-time emission of highly
relativistic material injects additional energy into the forward shock, 
flattening the decay slope \citep{kp97,rm00}.  
Alternatively, towards the end of the 
$\gamma$-ray emission, the central engine may inject material with a 
smooth distribution of (decreasing) Lorentz factors.  Slower moving
material will catch up with the forward shock when it has swept up
enough circum-burst material, resulting in a smooth injection of energy
at late times \citep{rm98,sm00}.  
While both models explain the flattening of the XRT
light curves, they provide different constraints on the nature of the
central engine.

In the first (long-lived central engine) model, 
the central engine's luminosity, $L(t)$, is characterized as
\begin{equation}
L(t) = L_{0} \left(\frac{t}{t_{0}}\right)^{-q}
\label{eqn:lum1}
\end{equation}
This results in the following spectral and temporal power-law indices for 
a constant-density medium:
\begin{eqnarray}
\label{eqn:c3a}
\nu < \nu_{\mathrm{m}}\;:\;\alpha & = & \frac{\displaystyle 5q - 8}
	{\displaystyle 6} \nonumber \\
	& = & (q-1) + \frac{(2+q)\beta}{2} \\
\label{eqn:c3b}
\nu_{\mathrm{m}} < \nu < \nu_{\mathrm{c}}\;:\;\alpha & = & 
	\frac{\displaystyle (2p-6) + (p+3)q}
	{\displaystyle 4} \nonumber \\
	& = & (q-1) + \frac{(2+q)\beta}{2} \\
\label{eqn:c3c}
\nu > \nu_{\mathrm{c}}\;:\;\alpha & = & \frac{\displaystyle (2p-4)+(p+2)q}
	{\displaystyle 4} \nonumber \\
   	& = & \frac{q-2}{2} + \frac{(2+q)\beta}{2} 
\end{eqnarray}
For a wind-like medium, the analogous results are
\begin{eqnarray}
\label{eqn:c4a}
\nu < \nu_{\mathrm{m}}\;:\;\alpha & = & \frac{\displaystyle q - 1}
	{\displaystyle 3} \nonumber \\
	& = & \frac{q}{2} + \frac{(2+q) \beta}{2} \\
\label{eqn:c4b}
\nu_{\mathrm{m}} < \nu < \nu_{\mathrm{c}}\;:\;\alpha & = & 
	\frac{\displaystyle (2p-2) + (p+1)q}
	{\displaystyle 4} \nonumber \\
	& = & \frac{q}{2} + \frac{(2+q) \beta}{2} \\
\label{eqn:c4c}
\nu > \nu_{\mathrm{c}}\;:\;\alpha & = & \frac{\displaystyle (2p-4) + (p+2)q}
	{\displaystyle 4} \nonumber \\
	& = & \frac{q-2}{2} + \frac{(2+q)\beta}{2}
\end{eqnarray}

The refreshed shock scenario is parameterized in terms of the amount of mass
ejected with Lorentz factor greater than $\gamma$: 
\begin{equation}
M(> \gamma) \propto \gamma^{-s}
\label{eqn:mass2}
\end{equation}
For the circum-burst profiles considered here, we can define a new variable, 
$\hat{q}$, such that we reproduce identical afterglow behavior to that
of Equations~\ref{eqn:c3a}-\ref{eqn:c3c} or Equations 
\ref{eqn:c4a}-\ref{eqn:c4c} 
by simply substituting $\hat{q}$ for $q$.  $\hat{q}$ is related
to the mass ejection parameter $s$ by the following equations \citep{zfd+06}:
\begin{eqnarray}
\hat{q} = & \frac{\displaystyle 10-2s}{\displaystyle 7+s} & \textrm{(ISM)} \\
\label{eqn:qsism}
	= & \frac{\displaystyle 4}{\displaystyle 3+s} & \textrm{(Wind)}
\label{eqn:qswind}
\end{eqnarray}

While the X-ray decay in Phase 2 is not as flat as that seen in
other \Swift\ bursts, the temporal and spectral decay indices are nonetheless
inconsistent with the standard afterglow model for $\nu(\mathrm{X}) >
\nu_{\mathrm{c}}$ (Equations \ref{eqn:c1d} and \ref{eqn:c2d}).  Furthermore,
the optical light curve shows a flattening during Phase 2, and is 
inconsistent with both the closure relations in either medium for 
$\nu_{\mathrm{m}} < \nu(\mathrm{opt}) < \nu_{\mathrm{c}}$ (Equations 
\ref{eqn:c1c} and \ref{eqn:c2c}).  We conclude we are therefore seeing a 
milder version of the energy injection phase present in many \Swift\ XRT
afterglows.

For the X-ray data in Phase 2, we find an acceptable fit for the energy
injection models only if $\nu(\mathrm{X}) > \nu_{\mathrm{c}}$.  This corresponds
to values of $q_{\mathrm{X}} = 0.66 \pm 0.08$ ($s_{\mathrm{X,ISM}} = 2.0 
\pm 0.3$,
$s_{\mathrm{X,Wind}} = 3.1 \pm 0.7$) 
and $p_{\mathrm{X}} = 2.4 \pm 0.2$.  
The optical data in Phase 2 are best fit with
a constant-density medium and $\nu_{\mathrm{m}} < \nu(\mathrm{opt}) < 
\nu_{\mathrm{c}}$: 
$q_{\mathrm{opt}} = 0.73 \pm 0.09$ ($s_{\mathrm{ISM,opt}} = 1.8 \pm 0.3$) and 
$p_{\mathrm{opt}} = 2.5 \pm 0.2$.  
Both the X-ray and the optical fall in the spectral regime
we would expect, providing further confidence in this interpretation.

A prediction of the energy injection hypothesis is a bright reverse shock
at early times most easily visible in the radio \citep{sm00}.  A reverse
shock nicely explains the rapid decline in flux at 8.5 GHz from 1--4 d
after the burst.  Furthermore, the transition from a spectrum
peaked around 8.5 GHz at $t_{\mathrm{BAT}} = 2.15$ d to an optically
thin radio spectrum at $t_{\mathrm{BAT}} = 4.11$ d can be understood
as the reverse shock peak frequency, $\nu_{\mathrm{m}}^{\mathrm{RS}}$,
passing through the radio.  Since $\nu_{\mathrm{m}}^{\mathrm{RS}} \approx
\nu_{\mathrm{m}}^{\mathrm{FS}} / \gamma^{2}$, this should occur
well before the forward shock peak frequency reaches the radio bands.

Distinguishing between the two theories to explain the energy injection
is quite difficult, as both models can be identically parameterized.
Progress in this area would require a large sample of bursts with detailed
contemporaneous X-ray and optical light curves.  If the refreshed
shocks are due to continued engine activity, they should be correlated with
the bright X-ray flares seen in some XRT afterglows.  On the other hand,
if the flat decay is caused by slow-moving ejecta, this behavior should
be more uniform from burst to burst.  Such an analysis is beyond the 
scope of this work.

\subsection{Burst Environment and Progenitor Models}
\label{sec:cbm}
We now turn our attention to the issue of the circum-burst medium.  As 
discussed earlier, the radial profile of the burst environment affects
the temporal decay below the cooling frequency (Equations 
\ref{eqn:t1a}-\ref{eqn:t1c}, \ref{eqn:t2a}-\ref{eqn:t2c}).  In particular,
the closure relationships (Equations \ref{eqn:c1a}-\ref{eqn:c1c},
\ref{eqn:c2a}-\ref{eqn:c2c}) are sufficiently different that we should be able
to distinguish between the competing models for a well-sampled event
like GRB\,050820A.

First we examine the X-ray data.  
As discussed previously (\S\ref{sec:inject}), the X-ray observations
in Phase 2 require invoking mild energy injection to explain the 
shallower-than-expected decay for $\nu(\mathrm{X}) > \nu_{\mathrm{c}}$. 
If we assume the X-ray spectral index does not change 
from Phase 2 to Phase 3, then we find $\alpha_{3,\mathrm{X}}$ and 
$\beta_{3,\mathrm{X}}$ satisfy the standard afterglow closure relation
for $\nu_{\mathrm{X}} > \nu_{\mathrm{c}}$ (Equations \ref{eqn:c1d} and
\ref{eqn:c2d}).  The corresponding values for the 
electron index are: $p_{\alpha_{3,\mathrm{X}}} = 2.3 \pm 0.1$, 
$p_{\beta_{3,\mathrm{X}}} = 2.40 \pm 0.08$.

Unlike the X-ray observations, the optical bands typically probe frequencies
below the cooling frequency, where the closure relations are
different for different circum-burst media (Equations \ref{eqn:c1c} and
\ref{eqn:c2c}).  We have shown already in \S\ref{sec:inject} that
the optical data in Phase 2 are better fit by a constant-density 
medium.  We find
that a constant-density medium 
is favored in the optical in Phase 1b$_{\mathrm{opt}}$ and Phase
3 as well.  The only closure relation satisfied in Phase 1b$_{\mathrm{opt}}$
is for an ISM with $\nu_{\mathrm{m}} < \nu(\mathrm{opt}) <
\nu_{\mathrm{c}}$ (Equation \ref{eqn:c1c}).  The resulting $p$-values are
$p_{\alpha_{1\mathrm{b,opt}}} = 2.29 \pm 0.02$, 
$p_{\beta_{1\mathrm{b,opt}}} = 2.1 \pm 0.1$.
We note that had we equated the BAT trigger time, $T_{\mathrm{BAT}}$, with
the onset of the burst ($T_{0}$), the temporal slope in 
Phase 1b$_{\mathrm{opt}}$ would not have been consistent with any closure
relation.

In Phase $3$, we cannot meaningfully constrain 
the optical spectral slope.  However, using the X-ray-to-optical spectral
slope in this phase, $\beta_{\mathrm{OX,3}} \approx 0.8$, we
conclude the optical data in this segment still fall below the cooling
frequency.  Based solely on the temporal decline then,
we can rule out a wind-like medium in this phase, 
The corresponding electron index
($p \approx 1.7 \pm 0.1$) would result in a divergent total energy.  While
this possibility has been addressed with more complicated electron
energy distributions (see e.g.~\citealt{dc01}), we consider this 
possibility unlikely. 

Taken together, the X-ray and optical data provide a consistent
picture of the forward shock expanding into a constant density medium.
The late-time ($t_{\mathrm{BAT}} > 1$ week)
radio observations, however, are inconsistent with this interpretation.  
For a constant-density medium, the peak flux density, 
$F_{\nu,\mathrm{max}}$, should remain constant
in time.  This would predict, if we have correctly interpreted
the optical peak as the forward shock (\S\ref{sec:early}), a similar peak 
($F_{\nu,\mathrm{max}} \approx 5$ mJy) in the radio at $t_{\mathrm{BAT}} 
\approx 7$ days
($\nu_{\mathrm{m}} \propto t^{-3/2}$).  
This is well above the VLA detection limit at this
epoch, yet we only measure $F_{\nu} \approx 100$ $\mu$Jy.  While the energy
injection phase will delay the arrival of $\nu_{\mathrm{m}}$ in the radio
($\nu_{\mathrm{m}}^{\mathrm{inject}} \propto t^{-3/2} t^{3(s-1)/2(7+s)}$), 
our radio
limits extend out to two months after the burst.  It would be very
difficult, if not impossible, to delay the peak this long.  Furthermore, during
the energy injection phase, the peak flux increases with time
($F_{\nu \mathrm{,max}}^{\mathrm{inject}} \propto t^{3(s-1)/(7+s)}$).
Thus we would expect to see rising emission earlier relative to the peak,
counteracting the delay of the peak radio flux.

One explanation for the lack of a bright, late-time radio afterglow is
an early jet break ($t \lesssim 1$ day), 
as was invoked for GRB\,990123 \citep{kfs+99}.
However, we find no evidence for a jet break in the optical or
X-ray light curves out to at least 17 days after the burst
(see \S\ref{sec:energy}).

Another possibility, invoked to explain the relatively low late-time
radio flux from GRB\,050904, is a high ambient density \citep{fck+06}.
In the case of GRB\,050904, it was argued that the large density raised the
self-absorption frequency, $\nu_{\mathrm{a}}$, above the radio observing
bands.  This greatly suppresses the radio flux, for the spectrum in this
regime is proportional to $\nu^{2}$ (Eqn.~\ref{eqn:nu1a}).  
There is no evidence in the radio data for an optically thick spectrum,
although spectral data is sparse at late times.  Furthermore,
broadband modeling of this event (\S\ref{sec:energy}) rules out
a high ambient density for typical values of the micro-physical
parameters $\epsilon_{\mathrm{e}}$ and $\epsilon_{\mathrm{B}}$.  We
therefore consider this explanation unlikely.
 
Alternatively, a natural explanation for the low radio flux at late times is a 
wind-like medium.  In a wind-like medium, 
the forward shock peak flux density declines in time as $F_{\nu,\mathrm{max}}
\propto t^{-3/2}$.  
The decreasing peak flux counteracts
the rising synchrotron emission, suppressing any late-time radio data.
This is of course inconsistent with our X-ray and optical data, 
which strongly favor a constant-density medium.  One can imagine a scenario 
in which the environment near the burst (the regime sampled predominantly
by the X-ray and optical data) is approximately constant in density, while
the outer regions (sampled by the radio at later times) have a wind-like
profile.  However, without a physical justification for such a density
profile, this remains little more than speculation.  The lack
of a bright radio afterglow remains a puzzling aspect of 
GRB\,050820A.

\begin{figure*}
\epsscale{1.0}
\plotone{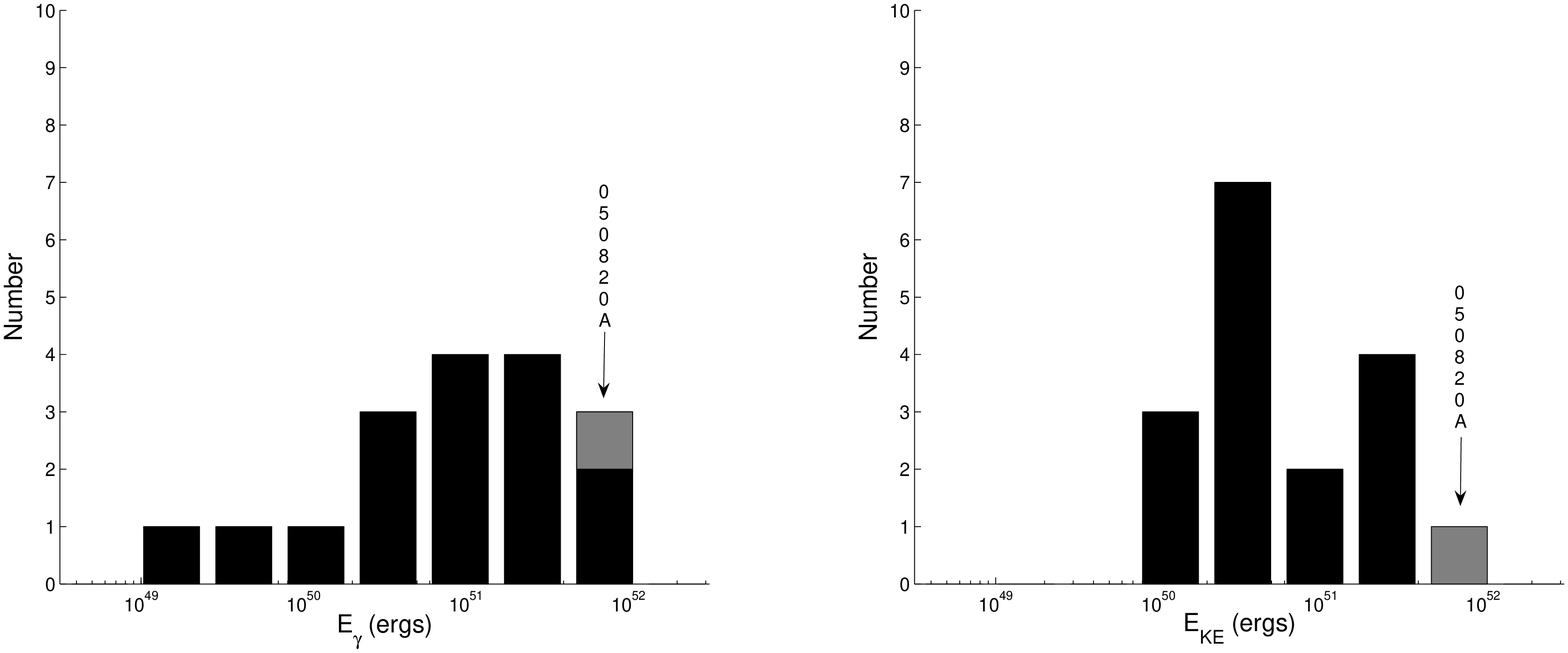}
\caption[Total Energy Release of GRB\,050820A]
        {Total energy release of GRB\,050820A.
        \textit{Left:} Collimation-corrected energy release in the prompt
        emission ($E_{\gamma}$) of a sample of cosmological
        GRBs, including GRB\,050820A.
        \textit{Right:} Collimation-corrected blast-wave energy
        ($E_{\mathrm{KE}}$)
        for the same sample.
        (References:
        \citealt{bkf04,yhs+03,pk02,bdf+01,clf04,skb+04,bkp+03}). }
\label{fig:energy}
\end{figure*}

\subsection{Geometry and Energetics}
\label{sec:energy}
Using the high-energy fluence derived from \KW\ satellite 
(\S\ref{sec:kw}), we calculate the total
isotropic energy release in the prompt emission was $E_{\gamma,\mathrm{iso}} = 
8.3^{+2.5}_{-1.1} \times 10^{53}$ ergs (assuming a redshift of $z = 
2.615$; \citealt{GCN.3833,GCN.3860}).  This makes GRB\,050820A one of
the most energetic events (in terms of $E_{\gamma,\mathrm{iso}}$) 
for which a redshift has been measured \citep{a06}.

However, only a fraction of the explosion energy is converted
into prompt emission via internal dissipation.  The rest remains in the 
kinetic energy of the outflow, powering the forward 
shock and hence the afterglow.
We can
estimate the kinetic energy of the afterglow ($E_{\mathrm{KE,iso}}$)
by examining the
the X-ray emission at $t_{\mathrm{BAT}} > 10$ hr \citep{fw01}.
At this point, the X-rays should be above the cooling frequency.  The
flux density is then independent of ambient 
density and only weakly dependent on
$\epsilon_{B}$.
A joint fit of the Phase 3 optical and X-ray data (after the energy
injection has stopped and the system has returned to adiabatic expansion)
constrains
the electron energy index: $p = 2.34 \pm 0.06$.  If we take typical values
for $\epsilon_{e}$ ($0.1 - 0.3$) and $\epsilon_{B}$ ($0.01 - 0.1$)
\citep{yhs+03}, we find
that $15 \lesssim E_{\mathrm{KE,iso,}52} \lesssim 100$. 

For an accurate accounting of the total
energy emitted by this event, however, we must determine the degree of
collimation of the emission.  We therefore examine all the temporal
breaks in the optical and X-ray light curves to determine which one (if any)
shows an achromatic steepening to the $t^{-p}$ decay expected from
a jet \citep{sph99}.  The only plausible candidate is the transition from
Phase 3 to 4 in the optical light curve.  The steepening here is achromatic
(i.e.~it is seen in all three \HST\ filters) and much too large to be
explained solely by the cooling frequency passing through the optical
bands (although this may have occured as well).  Any contribution from an 
underlying host galaxy would only further steepen the decay in Phase 4.

With only one observation, it is impossible to constrain the post-break
decay index.  Instead, we assume the post-break decay has a power-law 
index $\alpha = p \approx 2.34$ (see above).  We then find
$t_{\mathrm{jet,d}} = 18 \pm 2$ d.  This result is consistent with our
X-ray observations, which put a lower limit on the jet break 
time of $t_{\mathrm{jet,d}}
\ga 17$ d.

We note that the jet break time we have inferred for GRB\,050820A is 
extremely large.  In the host galaxy reference frame, the break
occurs at $t_{\mathrm{jet,d}}^{\mathrm{host}} \approx 5$, a factor of three
larger than any jet break seen in the pre-\Swift\ era \citep{zkk06}.  In
this respect, too, GRB\,050820A is a strong outlier.

To convert the jet break times to a range of opening angles, we use the
relation \citep{sph99}:
\begin{equation}
\label{eqn:thetatj}
\theta = 0.161 \left(\frac{t_{\mathrm{j}}}{1+z}\right)^{3/8} \left(\frac{
	n \eta_{\gamma}}{E_{\gamma\mathrm{,iso,52}}} \right)^{1/8}
\end{equation}
Here $\eta_{\gamma}$ 
is the fraction of the total energy converted to prompt $\gamma$-ray emission.
The only remaining unknown in Equation \ref{eqn:thetatj} is the ambient
density, $n$.  Using the ratio of the X-ray and optical data, as well as the
canonical values of $\epsilon_{\mathrm{e}}$ and $\epsilon_{\mathrm{B}}$, we 
find the density of the burst must be low: $n \le 1$ cm$^{-3}$.  
Afterglow modeling of the late-time optical and X-ray data (Phase 3,
after any continued energy injection has ceased and the shock
expands adiabatically) using
the technique of \citet{yhs+03} confirms this result: $n \lesssim 0.1$.  

Combining the above results, we find the opening angle is constrained
to fall between between
6.8$^{\circ} \lesssim \theta \lesssim 9.3^{\circ}$, corresponding to a 
beaming factor $f_{\mathrm{b}} \equiv 1 - \cos \theta \approx 10^{-2}$.  
While the opening angle is large for a long-soft burst, there are several
comparable events in the pre-\Swift\ sample \citep{zkk06}.
The total collimation-corrected energy emitted in $\gamma$-rays ($E_{\gamma}$)
from GRB\,050820A is therefore $7.5^{+6.7}_{-2.4} \times 10^{51}$ ergs.  The 
corresponding value for the blast-wave energy ($E_{\mathrm{KE}}$) is
$5.2^{+7.9}_{-4.1} \times 10^{51}$ ergs.

\citet{ggl04} have 
demonstrated an empirical relation between $E_{\gamma}$ and
the peak energy of the prompt emission spectrum in the GRB rest
frame ($E_{\mathrm{p}}^{\mathrm{rest}}$).  
GRB\,050820A is more energetic than any of the 37 bursts considered in their
sample (in terms of $E_{\gamma}$), 
and therefore proves an interesting test case for the so-called
Ghirlanda relation.  Using our calculated value of 
$E_{\gamma}$, Ghirlanda predicts $E_{\mathrm{p}}^{\mathrm{rest}} = 
2.0^{+2.5}_{-1.2}$ MeV.  This is marginally consistent with the actual value of 
$E_{\mathrm{p}}^{\mathrm{rest}} =
1.3^{+0.3}_{-0.2}$ MeV derived from the \KW\ dataset. 

In Figure \ref{fig:energy}, we plot a histogram of 
$E_{\gamma}$ and $E_{\mathrm{KE}}$ for the 
$\sim$ 15 long-soft cosmological bursts for 
which both quantities have been derived.  We have not included
the most nearby events (GRB\,980425, GRB\,031203, and GRB\,060218) in our
analysis, as these events released significantly less energy than
the typical cosmological GRB \citep{skn+06}. 
\citet{skb+04} have
shown that, with the exception of the most nearby events, the sum
of $E_{\gamma}$ and $E_{\mathrm{KE}}$ is clustered around 
$2 \times 10^{51}$ ergs.
GRB\,050820A is clearly an over-energetic exception, an order of
magnitude more energetic than this sample.
In fact, it would 
require the direct conversion of $\approx 
10^{-2}$  M$_{\odot}$ (with 100\% efficiency)
to release this much energy.

Finally, it is important to consider how robust our estimates of 
$E_{\mathrm{KE,iso}}$, $\theta$, and $n$ are 
given that the standard afterglow model fails
to explain the broadband behavior of GRB\,050820A.  We note that
the opening angle is relatively insensitive to both variables 
(Eqn.~\ref{eqn:thetatj}); factors of 
order unity will be greatly reduced by the $\frac{1}{8}$ exponent.
It is difficult to conceive of a long-soft 
GRB environment where the ambient density is
less than $10^{-2}$, and high densities would only increase the opening
angle and thus the energy release.  

\section{Conclusion}
\label{sec:conc}

GRB\,050820A joins a select sample of events with simultaneous observations
in the $\gamma$-ray and optical bands, and an even smaller group with 
contemporaneous X-ray observations as well.  Such events have led to 
fundamental advances in our understanding of GRBs, including the discovery
of a reverse-shock optical flash from GRB\,990123 \citep{abb+99} and 
possibly from GRB\,050904 \citep{bad+06}.  The early-time optical emission
from GRB\,041219A also showed a bright optical flash, but the rise was
correlated with an accompanying peak of $\gamma$-ray emission, suggesting
a common origin for the two components \citep{vww+05}.  

The early behavior of GRB\,050820A is unlike either of these events.
\citet{vww+06} have shown the contemporaneous optical emission is well
described as the sum of two components: one proportional to the prompt
$\gamma$-ray emission and one smoothly varying forward shock term.  While the
$\gamma$-ray component is important for $t_{\mathrm{BAT}} < 300$ s, the 
optical peak at $t_{\mathrm{BAT}} \approx 600$ s is dominated by emission
from the external shock region.
Furthermore, the post-peak decay rate is inconsistent with
reverse-shock emission.  Instead we interpret this as the forward
shock peak frequency passing through the optical bands.  This is not unlike 
what was seen in the optical for GRB\,060124 \citep{rcc+06}, although the 
time resolution in the prompt phase was much poorer than for this burst.
The contemporaneous optical light curves of GRB\,050319 \citep{qry+06,wvw+05} 
and
GRB\,050401 \citep{ryk+05} did not show this peak phase, but extrapolations
to late-times were consistent with the adiabatic expansion of a forward
shock.  
Another different behavior was seen in the early optical light curve of 
GRB\,050801, which showed an extended plateau phase correlated with the
X-ray emission, hinting at continued energy injection from a central engine
refreshing the external shocks \citep{rmy+06}.   

The contemporaneous X-ray emission, on the other hand, is the 
low-energy tail of the prompt emission.  This behavior was also seen
for GRB\,060124 \citep{rcc+06} and has been hinted at in the rapid
decline in early X-ray light curves attributed to high-latitude emission
\citep{lzo+06}, as well as the bright
X-ray flares seen in many XRT light curves \citep{brf+05}.  
It is clear then,
that, in marked contrast to the X-ray emission, contemporaneous optical
emission exhibits a large diversity in behavior.  Unfortunately the
physical mechanism behind this diversity remains to be explained.

The issue of burst geometry is a particularly interesting one in the \Swift\ 
era.  The steep
post-break decay slope, seen simultaneously in multiple filters, 
makes GRB\,050820A one of the most convincing
examples of a beamed event in the \Swift\ sample.   
The X-ray afterglow, however, is too faint at late times to
provide broadband confirmation.  In fact, very few \Swift\ bursts, including
those, like GRB\,050408 \citep{fpp+06},  
that have been followed for months, show 
signs of a jet break in the XRT light curve \citep{nkg+06}.

Typical jet breaks in pre-\Swift\ bursts occurred on time-scales of several
days \citep{zkk06}.  \citet{psf03} 
predicted that \Swift\ would detect bursts with
wider opening angles than previous missions due to the increased
sensitivity of the BAT.  However, 
not a single \Swift\ afterglow has shown a convincing
jet-break-like transition in 
multiple band-passes (candidates include GRB\,050525A and
GRB\,050801; see \citealt{pmb+06} and references therein for a more thorough
discussion).  While it may be that most jet breaks, 
like GRB\,050820A, occur at late times, beyond the sensitivity of the XRT and 
most ground-based facilities, this is nonetheless troubling.  On the one
hand, few if any X-ray jet breaks were seen in pre-\Swift\ bursts; all
collimation angles were determined from optical (and sometimes radio)
light curves.  Conversely, 
given the large number of well-sampled XRT light curves, and the fact that such
fundamental results for GRB cosmology as the Ghirlanda relation rest on our
picture of GRBs as a-spherical events, this is clearly a matter that merits
further investigation.

One consequence of the large opening angle associated with GRB\,050820A is a
correspondingly large burst and afterglow energy.  In fact, of all the bursts
compiled in the \citet{ggl+04} sample, GRB\,050820A has the largest prompt
energy release.  And unlike GRB\,990123 \citep{pk01}, this large
$\gamma$-ray energy was accompanied by a correspondingly large kinetic
energy imparted to the afterglow.  The only comparable event for which 
such energies could be determined was the high-redshift GRB\,050904,
which released a total energy of $\sim 10^{52}$ ergs \citep{tac+05,fck+06}.
Given the large $\gamma$-ray fluence,
similar events should have been easily detected by both \Swift\ and previous
GRB missions.  And given the bright optical afterglow and the late jet-break,
such events are strongly favored for ground-based follow-up (i.e.~redshift
determination).  The lack of a large sample of such events means they must
be relatively rare in the universe.

Like many other \Swift\ GRBs, the X-ray light curve of GRB\,050820A exhibits a 
phase of shallow decay incompatible with the standard forward shock model
\citep{nkg+06}.  GRB\,050820A is relatively
unique, however, in that this epoch is
also well-sampled in the optical.  The seemingly simultaneous breaks in
the optical light curve bolster the commonly-held belief that this phase
is caused by some form of refreshed shocks \citep{zfd+06}.  
Coupled with the large
gap between the precursor and the bulk of the prompt emission,
the late-time energy injection poses fundamental challenges to any central
engine model.

Finally, we return to the question of the radio afterglow.  Radio observations
typically probe low-Lorentz factor ejecta ($\Gamma \sim 2-3$) at large 
distances from the central engine ($r \sim 10^{17}$ cm).  The forward shock
peak frequency reaches the radio much later than the optical or X-ray
bands.  Thus, radio emission is usually 
visible at later times than optical or X-ray emission, and is 
well-suited to study afterglows when the emission is isotropic (i.e.~after
the jet break) or even in some cases when the ejecta has slowed to 
Newtonian expansion \citep{bkf04}.  
For this reason, late-time radio observations
are considered the most accurate method for model-independent calorimetry.  
For GRB\,050820A, this paradigm has broken down.  The burst had a
bright optical and X-ray afterglow, but weak emission in the radio.
It is hoped that further studies of such energetic GRBs in the \Swift\ era
will help to elucidate some of these puzzles.
 
\acknowledgments{We would like to thank the anonymous referee for helpful
comments on this manuscript.  
S.B.C.~and A.~M.~S.~are supported by the NASA Graduate
Student Research Program.  E.B.~is supported by NASA through Hubble
Fellowship grant HST-HF-01171.01 awarded by STScI, which is operated
by the Association of Universities for Research in Astronomy, Inc., for
NASA, under contract NAS5-26555.  A.G.~acknowledges support by NASA through
Hubble Fellowship grant HST-HF-01158.01 awarded by STScI.  The \KW\ 
experiment is supported by Russian Space Agency contract and RFBR grant
06-02-16070.  GRB research at Caltech is supported through NASA.
This publication has made use of data obtained from the \Swift\ interface
of the High-Energy Astrophysics Archive (HEASARC), provided by 
NASA's Goddard Space Flight Center.
This publication makes use of data products from the Two 
Micron All Sky Survey, which is a joint project of the University of 
Massachusetts and the Infrared Processing and Analysis Center/California 
Institute of Technology, funded by the National Aeronautics and Space 
Administration and the National Science Foundation.  The Digitized Sky Survey 
was produced at the Space Telescope Science Institute under U.S. Government 
grant NAG W-2166. The images of these surveys are based on photographic data 
obtained using the Oschin Schmidt Telescope on Palomar Mountain and the 
UK Schmidt Telescope. The plates were processed into the present compressed 
digital form with the permission of these institutions.}



\clearpage
\begin{deluxetable}{lrrrc}
  \tabletypesize{\footnotesize}
  \tablecaption{XRT Observations of GRB\,050820A}
  \tablecolumns{5}
  \tablewidth{0pc}
  \tablehead{\colhead{Mean Observation Date} &
	     \colhead{$t_{\mathrm{BAT}}$}
             & \colhead{Duration} & 
             \colhead{Spectral Index ($\beta$)} & 
	     \colhead{2--10 keV Flux} \\
             \colhead{(2005 UT)} & \colhead{(s)} & \colhead{(s)} &
	     & \colhead{($10^{-11}$ ergs cm$^{-2}$ s$^{-1}$)} 
            }
  \startdata
	Aug 20 06:36:25 & 92.0 & 10.0 & 0.90 & 
		$14.9 \pm 1.3$ \\
	Aug 20 06:36:35 & 102.0 & 10.0 & $\ldots$ &
		$12.2 \pm 1.2$ \\
	Aug 20 06:36:45 & 112.0 & 10.0 & $\ldots$ &
		$10.8 \pm 1.1$ \\
	Aug 20 06:36:55 & 122.0 & 10.0 & $\ldots$ &
		$9.8 \pm 1.1$ \\
	Aug 20 06:37:05 & 132.0 & 10.0 & $\ldots$ &
		$7.3 \pm 0.9$ \\
	Aug 20 06:37:15 & 142.0 & 10.0 & $\ldots$ &
		$8.1 \pm 1.0$ \\
	Aug 20 06:37:25 & 152.0 & 10.0 & $\ldots$ &
		$4.8 \pm 0.8$ \\
	Aug 20 06:37:35 & 162.0 & 10.0 & $\ldots$ &
		$4.5 \pm 0.8$ \\
	Aug 20 06:37:45 & 172.0 & 10.0 & $\ldots$ &
		$4.9 \pm 0.8$ \\
	Aug 20 06:37:55 & 182.0 & 10.0 & $\ldots$ &
		$2.5 \pm 0.6$ \\
	Aug 20 06:38:05 & 192.0 & 10.0 & $\ldots$ &
		$4.0 \pm 0.7$ \\
	Aug 20 06:38:15 & 202.0 & 10.0 & $\ldots$ &
		$2.3 \pm 0.8$ \\
	Aug 20 06:38:25 & 212.0 & 10.0 & $\ldots$ &
		$2.4 \pm 0.7$ \\
        \hline
	Aug 20 06:38:35 & 222.0 & 10.0 & -0.10  &
		$175.1 \pm 7.9$ \\
	Aug 20 06:38:45 & 232.0 & 10.0 & $\ldots$ &
		$567.0 \pm 14.2$ \\
	Aug 20 06:38:55 & 242.0 & 10.0 & $\ldots$ &
		$631.3 \pm 15.0$ \\
	Aug 20 06:39:05 & 252.0 & 10.0 & $\ldots$ &
		$629.7 \pm 26.5$ \\
        \hline
        Aug 20 07:56:43 & $4.910 \times 10^{3}$ & 250.0 & 
		1.20 & $3.0 \pm 0.2$ \\
        Aug 20 08:00:53 & $5.160 \times 10^{3}$ & 250.0 & 
		 $\ldots$ & $3.1 \pm 0.2$ \\
        Aug 20 08:05:03 & $5.410 \times 10^{3}$ & 250.0 & 
		 $\ldots$ & $3.2 \pm 0.2$ \\
        Aug 20 08:09:13 & $5.660 \times 10^{3}$ & 250.0 & 
		 $\ldots$ & $2.8 \pm 0.2$ \\
        Aug 20 08:13:23 & $5.901 \times 10^{3}$ & 250.0 & 
		 $\ldots$ & $2.6 \pm 0.2$ \\
        Aug 20 11:41:43 & $1.841 \times 10^{4}$ & $2.5 \times 10^{3}$ & 
		 $\ldots$ & $1.02 \pm 0.05$ \\
        Aug 20 12:23:23 & $2.091 \times 10^{4}$ & $2.5 \times 10^{3}$ & 
		 $\ldots$ & $0.9 \pm 0.1$ \\
        Aug 20 13:05:03 & $2.341 \times 10^{4}$ & $2.5 \times 10^{3}$ & 
		 $\ldots$ & $0.71 \pm 0.03$ \\
        Aug 20 14:28:23 & $2.841 \times 10^{4}$ & $2.5 \times 10^{3}$ & 
		 $\ldots$ & $0.58 \pm 0.03$ \\
        Aug 20 15:10:03 & $3.091 \times 10^{4}$ & $2.5 \times 10^{3}$ & 
		 $\ldots$ & $0.51 \pm 0.06$ \\
        Aug 20 15:51:43 & $3.341 \times 10^{4}$ & $2.5 \times 10^{3}$ & 
		 $\ldots$ & $0.52 \pm 0.04$ \\
        Aug 20 16:33:23 & $3.591 \times 10^{4}$ & $2.5 \times 10^{3}$ & 
		 $\ldots$ & $0.56 \pm 0.06$ \\
        Aug 20 17:15:03 & $3.841 \times 10^{4}$ & $2.5 \times 10^{3}$ & 
		 $\ldots$ & $0.47 \pm 0.07$ \\
        Aug 20 17:56:43 & $4.091 \times 10^{4}$ & $2.5 \times 10^{3}$ & 
		 $\ldots$ & $0.49 \pm 0.05$ \\
        Aug 20 19:20:03 & $4.951 \times 10^{4}$ & $2.5 \times 10^{3}$ & 
		 $\ldots$ & $0.39 \pm 0.08$ \\
        \hline
        Aug 21 15:48:01 & $1.196 \times 10^{5}$ & $1.0 \times 10^{4}$ & 
		 $\ldots$ & $0.18 \pm 0.03$ \\
        Aug 21 18:34:41 & $1.296 \times 10^{5}$ & $1.0 \times 10^{4}$ & 
		 $\ldots$ & $0.14 \pm 0.01$ \\
        Aug 21 21:21:21 & $1.396 \times 10^{5}$ & $1.0 \times 10^{4}$ & 
		 $\ldots$ & $0.13 \pm 0.02$ \\
        Aug 22 00:08:01 & $1.496 \times 10^{5}$ & $1.0 \times 10^{4}$ & 
		 $\ldots$ & $0.10 \pm 0.02$ \\
        Aug 22 02:54:41 & $1.596 \times 10^{5}$ & $1.0 \times 10^{4}$ & 
		 $\ldots$ & $0.07 \pm 0.01$ \\
        Aug 23 14:30:22 & $2.877 \times 10^{5}$ & $1.0 \times 10^{4}$ & 
		 $\ldots$ & $0.046 \pm 0.006$ \\
        Aug 23 17:17:02 & $2.977 \times 10^{5}$ & $1.0 \times 10^{4}$ & 
		 $\ldots$ & $0.050 \pm 0.008$ \\
        Aug 24 18:02:47 & $3.869 \times 10^{5}$ & $3.5 \times 10^{4}$ & 
		 $\ldots$ & $0.034 \pm 0.004$ \\
        Aug 25 03:46:07 & $4.219 \times 10^{5}$ & $3.5 \times 10^{4}$ & 
		 $\ldots$ & $0.05 \pm 0.01$ \\
        Aug 27 19:38:46 & $6.518 \times 10^{5}$ & $1.25 \times 10^{5}$ & 
		 $\ldots$ & $0.022 \pm 0.003$ \\
        Aug 29 06:22:06 & $7.768 \times 10^{5}$ & $3.5 \times 10^{4}$ & 
		 $\ldots$ & $0.014 \pm 0.002$ \\
        Sep 04 15:45:23 & $1.329 \times 10^{6}$ & $1.0 \times 10^{5}$ & 
		 $\ldots$ & $(6.8 \pm 2.5) \times 10^{-3}$ \\
        Sep 05 19:32:03 & $1.429 \times 10^{6}$ & $1.0 \times 10^{5}$ & 
		 $\ldots$ & $(4.2 \pm 1.5) \times 10^{-3}$ \\
        Sep 06 23:18:43 & $1.529 \times 10^{6}$ & $1.0 \times 10^{5}$ & 
		 $\ldots$ & $(8.8 \pm 3.3) \times 10^{-3}$ \\
  \enddata	
  \tablecomments{The four Phases of the X-ray light curve are delineated
	by horizontal lines (see \S\ref{sec:xraylc} for further details).
	We assumed the spectral index was constant in each Phase
	to convert count rates to the flux values shown here.  We also assumed
	the spectral index remained constant from Phase 2 to Phase 3.  All
	errors quoted are at the $1\sigma$ level.}
\label{tab:xrt}
\end{deluxetable}


\clearpage
\LongTables
\begin{deluxetable}{lrlrrcr}
  \tabletypesize{\footnotesize}
  \tablecaption{Optical Observations of GRB\,050820A}
  \tablecolumns{7}
  \tablewidth{0pc}
  \tablehead{\colhead{Mean Observation Date} & \colhead{$t_{\mathrm{BAT}}$} & 
             \colhead{Telescope} & \colhead{Filter} &
             \colhead{Exposure Time} & \colhead{Magnitude\tablenotemark{a}} & 
             \colhead{Reference} \\
	     \colhead{(2005 UT)} & \colhead{(s)} & & & \colhead{(s)} & 
	     \colhead{(Vega)} & 
            }
  \startdata 
	Aug 20 08:14:47 & $5.994 \times 10^{3}$ 
		& UVOT & $U$ & 693.3 & 18.12 $\pm$ 0.13 & * \\
	Aug 20 12:49:02 & $2.245 \times 10^{4}$ 
		& UVOT & $U$ & 899.8 & 19.11 $\pm$ 0.14 & * \\
	Aug 20 16:01:58 & $3.404 \times 10^{4}$
		 & UVOT & $U$ & 899.8 & 19.40 $\pm$ 0.14 & * \\
	Aug 20 19:14:54 & $4.560 \times 10^{4}$
		 & UVOT & $U$ & 899.8 & 19.66 $\pm$ 0.15 & * \\
	Aug 20 22:59:51 & $5.910 \times 10^{4}$ 
		 & UVOT & $U$ & 392.3 & 19.70 $\pm$ 0.18 & * \\
	\hline
	Aug 20 09:36:06 & $1.087 \times 10^{4}$
		 & UVOT & $B$ & 899.8 & 18.79 $\pm$ 0.16 & * \\
	Aug 20 13:04:10 & $2.336 \times 10^{4}$ 
		 & UVOT & $B$ & 899.8 & 19.24 $\pm$ 0.15 & * \\
	Aug 20 16:17:05 & $3.493 \times 10^{4}$
		 & UVOT & $B$ & 897.3 & 19.74 $\pm$ 0.16 & * \\
	Aug 20 19:30:02 & $4.651 \times 10^{4}$
		 & UVOT & $B$ & 899.8 & 19.87 $\pm$ 0.16 & * \\
	Aug 21 23:30:41 & $1.473 \times 10^{5}$ & RTT150 & $B$ & 1800.0 & 
		21.28 $\pm$ 0.06 & 1 \\
	Aug 22 22:58:17 & $2.318 \times 10^{5}$ & RTT150 & $B$ & 7860.0 & 
		22.05 $\pm$ 0.06 & 2 \\
	Aug 23 22:09:05 & $3.153 \times 10^{5}$ & RTT150 & $B$ & 5400.0 & 
		22.38 $\pm$ 0.08 & 2 \\
	\hline
	Aug 20 07:01:53 & $1.620 \times 10^{3}$
		 & P60 & $g$ & 120.0 & 16.27 $\pm$ 0.04 & * \\
	Aug 20 07:12:35 & $2.262 \times 10^{3}$
		 & P60 & $g$ & 120.0 & 16.65 $\pm$ 0.04 & * \\
	Aug 20 07:23:24 & $2.911 \times 10^{3}$
		 & P60 & $g$ & 120.0 & 16.92 $\pm$ 0.05 & * \\
	Aug 20 08:04:47 & $5.394 \times 10^{3}$
		 & P60 & $g$ & 360.0 & 17.59 $\pm$ 0.05 & * \\
	Aug 20 08:29:42 & $6.889 \times 10^{3}$
		 & P60 & $g$ & 360.0 & 17.87 $\pm$ 0.06 & * \\
	Aug 20 08:54:50 & $8.397 \times 10^{3}$
		 & P60 & $g$ & 360.0 & 18.11 $\pm$ 0.06 & * \\
	Aug 20 09:33:18 & $1.071 \times 10^{4}$
		 & P60 & $g$ & 720.0 & 18.29 $\pm$ 0.06 & * \\
	Aug 20 10:43:25 & $1.491 \times 10^{4}$
		 & P60 & $g$ & 720.0 & 18.36 $\pm$ 0.06 & * \\
	Aug 20 11:39:32 & $1.829 \times 10^{4}$
		 & P60 & $g$ & 720.0 & 18.59 $\pm$ 0.09 & * \\
	\hline
	Aug 20 06:36:58 & 125.0 & UVOT & $V$ & 89.0 & 18.18 $\pm$ 0.20 & 
		* \\
	Aug 20 07:52:58 & $4.685 \times 10^{3}$
		 & UVOT & $V$ & 99.8 & 17.20 $\pm$ 0.13 & * \\
	Aug 20 10:04:28 & $1.258 \times 10^{4}$
		 & UVOT & $V$ & 337.0 & 18.16 $\pm$ 0.12 & * \\
	Aug 20 11:12:34 & $1.666 \times 10^{4}$
		 & UVOT & $V$ & 899.8 & 18.39 $\pm$ 0.09 & * \\
	Aug 20 14:25:30 & $2.824 \times 10^{4}$
		 & UVOT & $V$ & 899.8 & 18.86 $\pm$ 0.11 & * \\
	Aug 20 17:38:26 & $3.981 \times 10^{4}$
		 & UVOT & $V$ & 899.8 & 19.35 $\pm$ 0.12 & * \\
	Aug 20 21:03:50 & $5.214 \times 10^{4}$
		 & UVOT & $V$ & 899.8 & 19.14 $\pm$ 0.11 & * \\
	Aug 21 16:09:34 & $1.209 \times 10^{5}$ & UVOT & $V$ & 899.8 & 
		20.24 $\pm$ 0.19 & * \\
	Aug 21 19:22:30 & $1.325 \times 10^{5}$ & UVOT & $V$ & 899.8 & 
		20.40 $\pm$ 0.22 & * \\
	\hline 
	Aug 20 06:38:49 & 236.0 & P60 & $R_{\mathrm{C}}$ & 60.0 & 
		15.39 $\pm$ 0.04 & * \\
	Aug 20 06:43:31 & 517.0 & P60 & $R_{\mathrm{C}}$ & 60.0 & 
		14.65 $\pm$ 0.02 & * \\
	Aug 20 06:48:25 & 812.0 & P60 & $R_{\mathrm{C}}$ & 60.0 & 
		15.06 $\pm$ 0.03 & * \\
	Aug 20 06:53:59 & $1.146 \times 10^{3}$ & P60 & 
		$R_{\mathrm{C}}$ & 120.0 & 15.42 $\pm$ 0.02 & * \\
	Aug 20 07:04:34 & $1.781 \times 10^{3}$ & P60 & 
		$R_{\mathrm{C}}$ & 120.0 & 16.05 $\pm$ 0.03 & * \\
	Aug 20 07:15:19 & $2.426 \times 10^{3}$ & P60 &  
		$R_{\mathrm{C}}$ & 120.0 & 16.39 $\pm$ 0.02 & * \\
	Aug 20 07:32:13 & $3.440 \times 10^{3}$ & P60 & 
		$R_{\mathrm{C}}$ & 360.0 & 16.77 $\pm$ 0.02 & * \\
	Aug 20 08:12:58 & $5.885 \times 10^{3}$ & P60 & 
		$R_{\mathrm{C}}$ & 360.0 & 17.32 $\pm$ 0.04 & * \\
	Aug 20 08:38:05 & $7.392 \times 10^{3}$ & P60 & 
		$R_{\mathrm{C}}$ & 360.0 & 17.48 $\pm$ 0.04 & * \\
	Aug 20 08:46:53 & $7.920 \times 10^{3}$ & PROMPT-5 & 
		$R_{\mathrm{C}}$ & 660.0 & 17.52 $\pm$ 0.09 & 3 \\
	Aug 20 09:03:19 & $8.906 \times 10^{3}$ & P60 & 
		$R_{\mathrm{C}}$ & 360.0 & 17.69 $\pm$ 0.04 & * \\
	Aug 20 09:28:50 & $1.044 \times 10^{4}$ & P60 & 
		$R_{\mathrm{C}}$ & 360.0 & 17.85 $\pm$ 0.04 & * \\
	Aug 20 10:00:44 & $1.235 \times 10^{4}$ & P60 & 
		$R_{\mathrm{C}}$ & 360.0 & 17.78 $\pm$ 0.04 & * \\
	Aug 20 10:38:59 & $1.465 \times 10^{4}$ & P60 & 
		$R_{\mathrm{C}}$ & 360.0 & 17.97 $\pm$ 0.04 & * \\
	Aug 20 11:05:54 & $1.626 \times 10^{4}$ & P60 & 
		$R_{\mathrm{C}}$ & 360.0 & 18.01 $\pm$ 0.04 & * \\
	Aug 20 11:33:14 & $1.790 \times 10^{4}$ & P60 & 
		$R_{\mathrm{C}}$ & 360.0 & 18.03 $\pm$ 0.04 & * \\
	Aug 20 12:02:31 & $1.966 \times 10^{4}$ & P60 & 
		$R_{\mathrm{C}}$ & 360.0 & 18.16 $\pm$ 0.05 & * \\
        Aug 21 00:37:53 & $6.498 \times 10^{4}$ & RTT150 & 
		$R_{\mathrm{C}}$ & 900.0 & 19.36 $\pm$ 0.01 & 1 \\
	Aug 21 04:16:53 & $7.812 \times 10^{4}$ & PROMPT-5 & 
		$R_{\mathrm{C}}$ & 5370.0 & 19.94 $\pm$ 0.31 & 3 \\
	Aug 22 00:00:05 & $1.491 \times 10^{5}$ & RTT150 & 
		$R_{\mathrm{C}}$ & 1800.0 & 20.26 $\pm$ 0.05 & 1 \\
	Aug 22 07:17:03 & $1.753 \times 10^{5}$ & P60 & $R_{\mathrm{C}}$ & 
		6840.0 & 20.51 $\pm$ 0.11 & * \\
	Aug 22 23:00:41 & $2.319 \times 10^{5}$ & RTT150 & $R_{\mathrm{C}}$ & 
		3900.0 & 20.90 $\pm$ 0.03 & 2 \\
	Aug 23 08:06:11 & $2.647 \times 10^{5}$ & P60 & $R_{\mathrm{C}}$ & 
		8400.0 & 20.89 $\pm$ 0.10 & * \\
	Aug 23 22:18:05 & $3.158 \times 10^{5}$ & RTT150 & $R_{\mathrm{C}}$ & 
		2700.0 & 21.18 $\pm$ 0.04 & 2 \\
	Aug 24 08:46:32 & $3.535 \times 10^{5}$ & P60 & $R_{\mathrm{C}}$ & 
		8400.0 & 21.22 $\pm$ 0.11 & * \\
	Aug 25 09:33:11 & $4.427 \times 10^{5}$ & P60 & $R_{\mathrm{C}}$ & 
		2880.0 & 21.34 $\pm$ 0.13 & * \\
	Aug 26 05:05:39 & $5.130 \times 10^{5}$ & HET & $R_{\mathrm{C}}$ & 
		600.0 & 21.57 $\pm$ 0.08 & * \\
	Aug 26 08:28:20 & $5.252 \times 10^{5}$ & P60 & $R_{\mathrm{C}}$ & 
		3600.0 & 21.64 $\pm$ 0.12 & * \\
	Aug 27 08:37:13 & $6.121 \times 10^{5}$ & P60 & $R_{\mathrm{C}}$ & 
		4800.0 & 21.80 $\pm$ 0.12 & * \\
	Aug 27 22:49:53 & $6.633 \times 10^{5}$ & RTT150 & $R_{\mathrm{C}}$ & 
		1500.0 & 22.02 $\pm$ 0.10 & 4 \\
	Sep 26 01:39:38 & $3.179 \times 10^{6}$ & HST & $F625W$ & 800.0 & 
		24.59 $\pm$ 0.08 & * \\
		&         &     & $R_{\mathrm{C}}$     &       & 24.55 $\pm$
		0.08 & * \\
	\hline
	Aug 20 06:40:21 & 328.0 & P60 & $I_{\mathrm{C}}$ & 60.0 & 
		14.91 $\pm$ 0.02 & * \\
	Aug 20 06:45:19 & 626.0 & P60 & $I_{\mathrm{C}}$ & 60.0 & 
		14.42 $\pm$ 0.01 & * \\
	Aug 20 06:50:06 & 913.0 & P60 & $I_{\mathrm{C}}$ & 60.0 & 
		14.78 $\pm$ 0.01 & * \\
	Aug 20 06:56:37 & $1.304 \times 10^{3}$ & P60 & $I_{\mathrm{C}}$ & 
		120.0 & 15.24 $\pm$ 0.02 & * \\
	Aug 20 07:07:15 & $1.942 \times 10^{3}$ & P60 & $I_{\mathrm{C}}$ & 
		120.0 & 15.74 $\pm$ 0.02 & * \\
	Aug 20 07:18:03 & $2.590 \times 10^{3}$ & P60 & $I_{\mathrm{C}}$ & 
		120.0 & 16.07 $\pm$ 0.02 & * \\
	Aug 20 07:40:11 & $3.918 \times 10^{3}$ & P60 & $I_{\mathrm{C}}$ & 
		360.0 & 16.54 $\pm$ 0.03 & * \\
	Aug 20 08:21:15 & $6.382 \times 10^{3}$ & P60 & $I_{\mathrm{C}}$ & 
		360.0 & 17.02 $\pm$ 0.05 & * \\
	Aug 20 08:46:26 & $7.893 \times 10^{3}$ & P60 & $I_{\mathrm{C}}$ & 
		360.0 & 17.22 $\pm$ 0.04 & * \\
	Aug 20 08:46:53 & $7.920 \times 10^{3}$ & PROMPT-3 & $I_{\mathrm{C}}$ 
		& 1560.0 & 17.31 $\pm$ 0.08 & 3 \\
	Aug 20 09:11:47 & $9.414 \times 10^{3}$ & P60 & $I_{\mathrm{C}}$ & 
		360.0 & 17.23 $\pm$ 0.05 & * \\
	Aug 20 09:37:27 & $1.095 \times 10^{4}$ & P60 & $I_{\mathrm{C}}$ & 
		360.0 & 17.34 $\pm$ 0.03 & * \\
	Aug 20 10:21:01 & $1.357 \times 10^{4}$ & P60 & $I_{\mathrm{C}}$ & 
		360.0 & 17.48 $\pm$ 0.04 & * \\
	Aug 20 10:47:55 & $1.518 \times 10^{4}$ & P60 & $I_{\mathrm{C}}$ & 
		360.0 & 17.68 $\pm$ 0.04 & * \\
	Aug 20 11:14:54 & $1.680 \times 10^{4}$ & P60 & $I_{\mathrm{C}}$ & 
		360.0 & 17.71 $\pm$ 0.05 & * \\
	Aug 20 11:42:22 & $1.845 \times 10^{4}$ & P60 & $I_{\mathrm{C}}$ & 
		360.0 & 17.72 $\pm$ 0.04 & * \\
	Aug 21 04:04:53 & $7.740 \times 10^{4}$ & PROMPT-3 & $I_{\mathrm{C}}$ 
		& 5440.0 & 18.33 $\pm$ 0.11 & 3 \\
	Aug 22 00:07:53 & $1.496 \times 10^{5}$ & RTT150 & $I_{\mathrm{C}}$ & 
		1800.0 & 19.74 $\pm$ 0.07 & 1 \\
	Aug 22 07:25:05 & $1.758 \times 10^{5}$ & P60 & $I_{\mathrm{C}}$ & 
		6840.0 & 19.97 $\pm$ 0.12 & * \\
	Aug 22 23:06:41 & $2.323 \times 10^{5}$ & RTT150 & $I_{\mathrm{C}}$ & 
		3900.0 & 20.37 $\pm$ 0.05 & 2 \\
	Aug 23 08:22:42 & $2.657 \times 10^{5}$ & P60 & $I_{\mathrm{C}}$ & 
		7560.0 & 20.48 $\pm$ 0.12 & * \\
	Aug 23 22:24:05 & $3.162 \times 10^{5}$ & RTT150 & $I_{\mathrm{C}}$ & 
		2700.0 & 20.78 $\pm$ 0.09 & 2 \\
        Aug 24 09:11:48 & $3.550 \times 10^{5}$ & P60 & $I_{\mathrm{C}}$ & 
		8400.0 & 20.59 $\pm$ 0.11 & * \\
	Aug 25 10:44:39 & $4.470 \times 10^{5}$ & P60 & $I_{\mathrm{C}}$ & 
		2400.0 & 21.02 $\pm$ 0.15 & * \\
	Aug 26 04:49:33 & $5.121 \times 10^{5}$ & HET & $I_{\mathrm{C}}$ & 
		1200.0 & 21.25 $\pm$ 0.10 & * \\
	Aug 26 09:27:24 & $5.288 \times 10^{5}$ & P60 & $I_{\mathrm{C}}$ & 
		4800.0 & 21.17 $\pm$ 0.14 & * \\
	Aug 27 08:34:02 & $6.120 \times 10^{5}$ & P60 & $I_{\mathrm{C}}$ & 
		4440.0 & 21.30 $\pm$ 0.13 & * \\
	Sep 26 02:01:58 & $3.180 \times 10^{6}$ & HST & $F775W$ & 800.0 & 
		24.32 $\pm$ 0.09 & * \\
		& & & $I_{\mathrm{C}}$ & & 24.27 $\pm$ 0.09 & * \\
	\hline
        Aug 20 06:41:50 & 417.0& P60 & $z'$ & 60.0 & 13.93 $\pm$ 0.11 &
		* \\
	Aug 20 06:46:45 & 712.0 & P60 & $z'$ & 60.0 & 14.27 $\pm$ 0.14 &
		* \\
	Aug 20 06:51:50 & $1.017 \times 10^{3}$ & P60 & $z'$ & 60.0 & 
		14.62 $\pm$ 0.21 & * \\
	Aug 20 06:59:16 & $1.463 \times 10^{3}$ & P60 & $z'$ & 120.0 & 
		15.02 $\pm$ 0.13 & * \\
	Aug 20 07:09:56 & $2.103 \times 10^{3}$ & P60 & $z'$ & 120.0 & 
		15.49 $\pm$ 0.14 & * \\
	Aug 20 07:20:45 & $2.752 \times 10^{3}$ & P60 & $z'$ & 120.0 & 
		15.96 $\pm$ 0.20 & * \\
	Sep 26 03:06:14 & $3.184 \times 10^{6}$ & HST & $F850LP$ & 1600.0 & 
		24.09 $\pm$ 0.09 & * 
  \enddata
  \tablenotetext{a}{Errors quoted are 1$\sigma$ photometric and instrumental
	errors summed in quadrature.  Galactic extinction ($E(B-V) = 0.044$;
	\citealt{sfd98}) has been incorporated in the reported magnitudes.}
  \tablerefs{* = this work; 1 = \citet{GCN.3853}; 2 = \citet{GCN.3864};
	3 = \citet{GCN.3863}; 4 = \citet{GCN.3896}.}
\label{tab:optical}
\end{deluxetable}

\end{document}